\documentclass[trackchanges]{aastex701}
\usepackage{float}
\usepackage[inkscapeformat=png]{svg}
\usepackage{xspace}
\usepackage[normalem]{ulem}
\usepackage{amsmath}
\usepackage[table]{xcolor}
\usepackage{colortbl} % Optional but useful
\usepackage{graphicx}
\usepackage{subcaption}
\usepackage{makecell} % add this in your preamble\
\usepackage{multirow}

\begin{document}

\title{Solar Extreme Ultraviolet Spectrograph and High-energy Imager (SEUSHI): Design, Development, and Pre-Flight Calibration}

\author[orcid=0000-0001-5047-5133,sname='Telikicherla']{Anant Telikicherla}
\altaffiliation{Ann and H.J. Smead Department of Aerospace Engineering Sciences, University of Colorado at Boulder}
\affiliation{Laboratory for Atmospheric and Space Physics, University of Colorado at Boulder, 3665 Discovery Dr., Boulder, CO 80303}
\email[show]{anant.telikicherla@lasp.colorado.edu}  

\author[orcid=0000-0002-2308-6797, sname='Woods']{Thomas N. Woods} 
%\altaffiliation{Las Campanas Observatory}
\affiliation{Laboratory for Atmospheric and Space Physics, University of Colorado at Boulder, 3665 Discovery Dr., Boulder, CO 80303}
\email{tom.woods@lasp.colorado.edu}

\author[orcid=0009-0004-7222-1512, sname='Crotser']{Dave Crotser}
\affiliation{Laboratory for Atmospheric and Space Physics, University of Colorado at Boulder, 3665 Discovery Dr., Boulder, CO 80303}
\email{Dave.Crotser@lasp.colorado.edu}

\author[orcid=0000-0002-1426-6913, sname='Schwab']{Bennet D. Schwab}
\affiliation{Space Sciences Laboratory, University of California Berkeley, 7 Gauss Way, Berkeley, CA 94720}
\email{bennetschwab@berkeley.edu}

\author[orcid=0000-0002-3440-2492, sname='Sewell']{Robert H. Sewell}
\affiliation{Laboratory for Atmospheric and Space Physics, University of Colorado at Boulder, 3665 Discovery Dr., Boulder, CO 80303}
\email{Robert.Sewell@lasp.colorado.edu}

\author[orcid=0000-0003-3210-5563, sname='ZagorecMarks']{Wyatt ZagorecMarks}
\affiliation{Laboratory for Atmospheric and Space Physics, University of Colorado at Boulder, 3665 Discovery Dr., Boulder, CO 80303}
\email{wyatt.zagorecMarks@lasp.colorado.edu}

\author[orcid=0000-0002-4546-2394, sname='Sims']{Alan Sims}
\affiliation{Laboratory for Atmospheric and Space Physics, University of Colorado at Boulder, 3665 Discovery Dr., Boulder, CO 80303}
\email{alan.sims@lasp.colorado.edu}

\author[orcid=0000-0001-5533-5498, sname='Jones']{Andrew R. Jones}
\affiliation{Laboratory for Atmospheric and Space Physics, University of Colorado at Boulder, 3665 Discovery Dr., Boulder, CO 80303}
\email{andrew.jones@lasp.colorado.edu}

\author[orcid=0000-0002-3783-5509, sname='Mason']{James P. Mason}
\affiliation{Johns Hopkins University Applied Physics Laboratory, 11100 Johns Hopkins Rd, Laurel, MD 20723}
\email{James.Mason@jhuapl.edu}

\author[orcid=0000-0003-4372-7405, sname='Chamberlin']{Philip Chamberlin}
\affiliation{Laboratory for Atmospheric and Space Physics, University of Colorado at Boulder, 3665 Discovery Dr., Boulder, CO 80303}
\email{Phil.Chamberlin@lasp.colorado.edu}

\begin{abstract}
Understanding the initiation of solar flares and coronal mass ejections (CMEs) is essential for improving forecasts of extreme space weather. Soft X-ray (SXR) and Extreme Ultraviolet (EUV) observations provide critical diagnostics of the highly dynamic solar corona; however, significant measurement gaps persist despite decades of observations. The Solar Extreme Ultraviolet Spectrograph and High-energy Imager (SEUSHI) aims to address these gaps by combining multi-pinhole SXR imaging with grazing-incidence EUV spectroscopy on a shared camera. SEUSHI delivers spatially-resolved temperature and emission measure maps at 1 arcminute resolution and 5 second cadence to identify Hot Onset Precursor Events (HOPEs), which provide early alerts of flares. Additionally, high-cadence (100 Hz) readouts of selected image rows enable photon-counting spectroscopy over 1.1–6.8 keV with $\mathrm{\approx}$0.08 keV energy resolution, to investigate elemental abundance evolution in active regions, a key diagnostic of coronal heating. SEUSHI also provides high-resolution ($\mathrm{\approx}$0.2 nm) EUV spectra measurements across the 16.1–33.8 nm range at 5 second cadence for studies of coronal dimming and generation of early alerts for CMEs. SEUSHI is designed with low power, mass, and volume requirements, making it suitable for small satellite platforms. A technology demonstration version of SEUSHI is currently under development for flight aboard the Solar Dynamics Observatory Extreme Ultraviolet Variability Experiment calibration sounding rocket. This paper presents the instrument design, development, and calibration. Successful demonstration on the sounding rocket platform is an important step towards the opportunity to fly SEUSHI on future satellite missions and thus to improve space weather operations.
\end{abstract}


\section{Introduction} \label{sec:intro}
The Sun’s outer atmosphere, the corona, is highly dynamic and produces solar eruptive events that affect the near-Earth space environment. These events include solar flares, coronal mass ejections (CMEs), and solar energetic particle (SEP) storms (for example see, \cite{kawabata_statistical_2018} and references therein). Solar Eruptive Events can degrade or disrupt communication and navigation systems by affecting the ionosphere and atmosphere \citep{curto_geomagnetic_2020}. Additionally, high energy particles can damage satellite infrastructure and pose a threat to astronauts \citep{chancellor_space_2014}. Understanding the physical mechanisms driving solar eruptions is essential for improving predictions of extreme space weather. Solar flares, the first sign of a space weather event, emit electromagnetic radiation across a wide range of wavelengths, from X-rays to radio waves. Emission in the Extreme Ultraviolet Bands (EUV: 0.01 to 0.124 keV or 10 to 121 nm) and Soft X-Ray band (SXR: 0.124 to 12.24 keV or 0.1 to 10 nm) can increase by several orders of magnitude during flares on timescales of a few minutes. These emissions provide key diagnostics about the heating processes involved during the initiation of solar eruptive events. 

Current state-of-the-art SXR observations include measurements from the Geostationary Operational Environmental Satellite X-Ray Sensor (GOES-XRS), which provides full-disk–integrated solar irradiance in two broad energy bands (called XRS-A and XRS-B) \citep{woods_goes-r_2024}. In addition, spectrometers such as the Miniature X-ray Solar Spectrometer (MinXSS) \citep{mason_minxss-2_2020}, the Dual Aperture X-ray Solar Spectrometer (DAXSS) \citep{woods_first_2023}, the Solar Low Energy X-ray Spectrometer (SoLEXS) aboard \textit{Aditya-L1} \citep{sankarasubramanian_solar_2025}, and the X-ray Solar Monitor (XSM) \citep{mithun_solar_2020} aboard \textit{Chandrayaan} provide high spectral resolution measurements across the SXR band. However, these observations are disk-integrated and therefore lack spatial information about the emitting regions. As solar flares are highly localized phenomena, the use of full-disk–integrated measurements introduces uncertainties, particularly related to background subtraction and source confusion. Consequently, there is a strong need for spatially-resolved SXR spectroscopy to better constrain localized flare physics. The Spectrometer/Telescope for Imaging X-rays (STIX) aboard the \textit{Solar Orbiter} mission \citep{krucker_spectrometertelescope_2020} provides both imaging and spectroscopy, but its spectral resolution in the soft X-ray regime is lower (1 keV at 6 keV) than that of dedicated spectrometers such as MinXSS, DAXSS, SoLEXS, and XSM (approximately, 0.05 keV at 1 keV). Previous instruments, such as the Soft X-ray Telescope (SXT) aboard \textit{Yohkoh} \citep{lw_-orbit_2016} and the X-Ray Telescope (XRT) aboard \textit{Hinode} \citep{golub_x-ray_2008}, provide high-resolution broadband SXR images but rely on filter wheels to sequentially sample different passbands. As a result, they cannot provide simultaneous measurements across multiple wavelengths and techniques such as the filter-ratio method to determine plasma properties (temperature and emission measure) cannot be applied during the flare onset and impulsive phases, when rapid temporal evolution is most pronounced. 

In the EUV, there are multiple imagers currently operational including the Atmospheric Imaging Assembly (AIA) onboard the Solar Dynamics Observatory (SDO) \citep{lemen_atmospheric_2012}, the Solar Ultraviolet Imager (SUVI) on the GOES-R series \citep{darnel_goes-r_2022}, and the Extreme Ultraviolet Imager (EUI) on Solar Orbiter \citep{rochus_solar_2020}. Additionally, there are multiple irradiance instruments, including Extreme Ultraviolet Variability Experiment (EVE) on SDO \citep{woods_extreme_2012}, Extreme Ultraviolet and X-ray Irradiance Sensors (EXIS) on GOES-R \citep{machol_goes-r_2020}, Solar EUV Monitor (SEM) on SOHO \citep{judge_first_1998}, Large Yield RAdiometer (LYRA) on PROBA-2 \citep{hochedez_lyra_2006}, and Extreme Ultraviolet Monitor (EUVM) on MAVEN \citep{eparvier_solar_2015} that provide broadband or spectral irradiance measurements. Lastly, instruments providing imaging spectroscopy include Spectral Imaging of the Coronal Environment (SPICE) on Solar Orbiter \citep{anderson_solar_2020} and the EUV Imaging Spectrometer (EIS) on Hinode \citep{culhane_euv_2007}. However, due to a CCD electronics capacitor failure in 2014 that permanently disabled the MEGS-A channel of EVE, there exists a gap in full-disk spectral irradiance measurements in the critical 6–33 nm wavelength range, previously covered at high spectral resolution and cadence \citep{woods_extreme_2012}. The EUVS-A channel of EXIS instrument, has a wavelength range of 25-31 nm and partially covers this band, but at low spectral resolution (0.6 nm) at particular lines of interest \citep{machol_goes-r_2020}.

The Solar Extreme Ultraviolet Spectrograph and High-energy Imager (SEUSHI) is being developed to address these gaps by combining multi-pinhole SXR imaging with grazing-incidence EUV spectroscopy on a shared Complementary Metal Oxide Semiconductor (CMOS) sensor. SEUSHI acquires images in multiple SXR passbands, enabling the derivation of plasma temperature and emission measure using the filter-ratio method. This method has been used extensively with GOES-XRS irradiance data \cite{white_updated_2005}, but without accounting for spatial variation. In contrast, SEUSHI observes multiple passbands simultaneously, enabling early detection of solar flares and investigation of energy release processes associated with the Hot Onset Precursor Event (HOPE) \citep{hudson_hot_2021, battaglia_existence_2023, dasilva_statistical_2023, telikicherla_investigating_2024, telikicherla_improving_2025}. Furthermore, by performing fast readout of selected sensor rows, SEUSHI operates in a photon-counting mode to obtain spatially resolved SXR spectra, enabling imaging spectroscopy in a compact instrument design. In addition to the SXR imaging spectrometer the SEUSHI instrument consists of a EUV spectrograph. The EUV spectrograph provides the capability of measuring coronal dimming, which provides an early alert for CMEs directed towards the Earth. Coronal dimming, which is a signature of rapid plasma evacuation from CME footpoints, scales with both the mass and kinetic energy of the eruption, allowing the magnitude and initial velocity of CMEs to be estimated early from the dimming area, rate, and/or depth (for example see \citep{mason_sdoeve_2019} and references therein). Additionally, another objective of the EUV spectrograph is to provide a low mass, cost and power alternative to the existing SDO-EVE Multiple EUV Grating Spectrograph A channel (MEGS-A) \citep{woods_extreme_2012}. Since the underflight rocket experiment instrument has been reflown for over a decade having an alternate instrument for underflight calibration is important to ensure redundancy in case the main instrument has any issues, and that risk was the original motivation for the development of a compact version of EVE Multiple EUV Grating Spectrograph (MEGS), which then evolved into the SEUSHI configuration. 

The SEUSHI instrument is being developed for launch onboard the Solar Dynamics Observatory Extreme-ultraviolet Experiment (SDO-EVE) \citep{woods_extreme_2012} under-flight calibration rocket mission in August 2026. Additional details of the science and space weather measurement capabilities of the instrument are described in the next section. During its sounding rocket flight, the SEUSHI instrument will collect about 10 minutes of scientific data, and a separate publication will describe the first-light results and scientific analysis of the same. This paper describes the design, development, and pre-flight calibration of the SEUSHI instrument, and is organized as follows: Section \ref{sec:stm} describes the science and space weather traceability matrix of the instrument, Section \ref{sec:instoverview} provides the instrument overview and explains the different instrument components, Section \ref{sec:design} provides a detailed discussion of the different design parameters of the instrument, Section \ref{sec:allign_cal} describes the alignment and calibration efforts, and lastly Section \ref{sec:Future_Work} describes the future improvements that can be performed to improve instrument performance for deployment on future small satellite missions. The details of the SEUSHI design optimization and signal analysis are provided in Appendix \ref{sec:appendixA}.

\section{Instrument Measurement Requirements} \label{sec:stm}
In order to clearly define the measurement requirements of the SEUSHI instrument, a science traceability matrix (STM) was created that flows the science/space weather goals to measurement requirements. The first column of the STM shows describes the fundamental science (solar physics) research goal and and the second column describes the space weather objective focused on flare and CME forecasting important for space weather operations. The overall motivation of the instrument stems from the Heliophysics Decadal Survey that has a significant focus on improving space weather alerts and forecasts as described in Priority Science Goal 3.d, Priority Science Goal 2.b, and Space Weather Goal E.3.1 \citep{decadal_correct}. Table \ref{tab:stm} shows the science/space weather traceability matrix of the instrument with calls to specific decadal survey goals in the table header. The instrument has three main science questions and space weather objectives, all focused towards solar eruptive events. 

The first science questions involves improving the understanding of plasma parameters during the HOPE phase in solar flares. The HOPE phenomenon, which has been the subject of many recent studies \citep{hudson_hot_2021, dasilva_statistical_2023, battaglia_existence_2023, telikicherla_investigating_2024, telikicherla_improving_2025}, has shown that most flares exhibit elevated plasma temperature (10-15 MK) and emission measure (10$^{46}$ - 10$^{50}$ cm$^{-3}$) prior to the impulsive phase of the flare by several minutes. Most studies to date have utilized full-disk integrated measurements, which introduces approximations because of background subtraction. The SEUSHI instrument aims to create spatially resolved temperature and emission measure maps of the sun which can then be used to apply the HOPE nowcasting algorithms to generate early flare alerts at the pixel level \citep{hudson_anticipating_2025, telikicherla_improving_2025}. Consequently, the first objective involves generating real-time early flare nowcast alerts at the pixel level utilizing the HOPE effect in solar flares. This would improve existing flare alerts and also predict the flare location on the limb. Understanding the flare location on the solar limb is beneficial for space weather, particularly to know if an associated CME or SEP event is earth directed or not. 

The second objective of the SEUSHI instrument is to understand evolution of active region plasma parameters including temperature, emission measure, and elemental abundance factors. Many previous studies (e.g., \cite{cargill_modelling_2015, reale_coronal_2014}) have investigated the coronal temperature and emission measure for active regions, however the active region abundance variations remain poorly understood (e.g., \cite{del_zanna_elemental_2014}). This is primarily because most SXR spectrum measurements to data (e.g., \cite{mason_minxss-2_2020, woods_first_2023}) use disk-integrated measurements, where it is difficult to isolate active region spectra. The solar SXR spectra in the 1.5-4 keV range is optimal for studying the abundance variations of several first-ionization-potential (FIP) elements (Fe, Mg, Si) that are known to vary with different coronal heating processes (e.g., \cite{laming_fip_2015}). The SEUSHI instrument will obtain spatially resolved SXR spectra by fast readout of the SEUSHI sensor so as to detect individual SXR photon's energy and build up a SXR spectra for the brighter active regions. By doing this, the SEUSHI instrument aims to study how the plasma parameters of active regions evolve overtime, which could provide more insight into active region heating phenomena and also indicate subsequent solar eruptive events. 

\begin{deluxetable*}{l l l l l c c}
\tablecaption{Instrument Science/Space Weather Traceability Matrix. \label{tab:stm} \\
\textbf{Instrument Goal:} Near-real-time detection and characterization of solar flares and CMEs using coordinated EUV and SXR observations. \\
\textbf{Heliophysics Decadal Survey Goals:} Priority Science Goal 3.d\tablenotemark{a}; Priority Science Goal 2.b\tablenotemark{b}; Space Weather E.3.1 Goal 1\tablenotemark{c}
}
\tablewidth{0pt}
\tabletypesize{\footnotesize}
\renewcommand{\arraystretch}{1.1}
\setlength{\tabcolsep}{3pt}
\tablehead{
\colhead{Science Question} &
\colhead{Space Weather Objective} &
\colhead{Measurement Description} &
\colhead{Physical Parameters} &
\colhead{Measurement} &
\colhead{Requirement} &
\colhead{Performance}
}
\startdata
\multirow{7}{*}{\makecell[l]{What are the coronal\\ plasma
properties during\\
the onset and early\\evolution of solar flares?}} &
\multirow{7}{*}{\makecell[c]{Provide accurate near\\
 real-time nowcast alerts\\ for solar flares}} &
\multirow{7}{*}{\makecell[c]{Utilize Hot Onset Precursor\\
Event (HOPE) plasma\\
parameter signatures}} &
\multirow{7}{*}{\makecell[c]{Flaring plasma temperature\\
(6--30 MK);\\
Emission measure\\
($10^{46}$--$10^{50}$ cm$^{-3}$)}} &
Field of View & $\geq 1\,R_{\odot}$\textsuperscript{1} & $\mathrm{1.3\,R_{\odot}}$\textsuperscript{1} \\
& & & & Spatial Resolution & $\leq 60''$ & $57''$ \\
& & & & Energy Range & 1.5--4 keV & 1.8--6.8 keV \\ %1.1--6.9 keV, 
& & & & Spectral Bands & $\geq 2$ & 2 \\
& & & & Time Cadence & $\leq 60$ seconds & 5 seconds \\
& & & & SNR (for M1 Flare) & $\mathrm{\geq 10}$ & $\approx$50\\
& & & & Accuracy & $\leq 20\%$ & $\approx$10\% \\
\hline
\multirow{7}{*}{\makecell[l]{How do active region\\
thermal structure and\\
composition evolve prior\\
to and between solar\\
eruptive events?}} &
\multirow{7}{*}{\makecell[c]{Enhance understanding\\
of solar active regions to\\
improve forecasting\\methods}} &
\multirow{7}{*}{\makecell[c]{Utilize spatially resolved\\
SXR spectra to track active\\
region evolution}} &
\multirow{7}{*}{\makecell[c]{Active region plasma temperature\\
(4--6 MK);\\
Emission measure\\($10^{44}$--$10^{46}$ cm$^{-3}$);\\
Elemental abundances\\
(1--4$\times$ photospheric)}} &
Field of View & $>4'$ & $\mathrm{1.3\,R_{\odot}}$\textsuperscript{1} \\
& & & & Spatial Resolution & $\leq 60''$ & $57''$ \\
& & & & Energy Range & 1.5--4 keV & 1.1--6.9 keV \\
& & & & Energy Resolution\textsuperscript{2} & 0.1 keV & 0.08 keV\\
& & & & Time Cadence & $\leq 1$ hr & 10 minutes \\
& & & & SNR (at 2 keV)  & $\geq 10$ & $\approx$38.9\\
& & & & Accuracy & $\leq 20\%$ & $\approx$10\% \\
\hline
\multirow{5}{*}{\makecell[l]{What are the coronal\\plasma properties during\\the onset and early\\ evolution of Coronal\\ Mass Ejections?}} &
\multirow{5}{*}{\makecell[c]{Provide accurate near\\
real-time forecast alerts\\
for CMEs}} &
\multirow{5}{*}{\makecell[c]{Utilize coronal dimming observed\\
in EUV spectra and flare/CME\\
location from SXR images}} &
\multirow{5}{*}{\makecell[c]{CME mass ($1.2\cdot10^{13}$--$7.8\cdot10^{15}$g);\\
CME velocity (110 -- 2080 km/s);\\
CME direction (Visible Hemisphere)}} &
Spectral Range & 17--31 nm & 16.1--33.8 nm \\
& & & & Spectral Resolution & $<0.6$ nm & 0.2 nm \\
& & & & Time Cadence & $\leq 60$ seconds & 5 seconds \\
& & & & SNR (at 193$\mathrm{\AA}$) & $\geq 400$ & $\approx$200 \\
& & & & Accuracy & $\leq 20\%$ & $\approx$10\% \\
\enddata
\tablenotetext{a}{Improve methods for forecasting and nowcasting solar eruptions at the Sun and their impacts on interplanetary radiation environments.}
\tablenotetext{b}{Determine the role quasi-steady processes play in heating the solar corona and accelerating the solar wind and suprathermal particles.}
\tablenotetext{c}{Develop an accurate and reliable 12-hour lead-time probabilistic $>$M1 solar eruption forecast and associated SEP event forecast with 6-hour lead time.}
\tablenotetext{1}{$\mathrm{\,R_{\odot}}$ denotes the radius of the sun, which is 32 arcminutes as viewed from Earth.}
\tablenotetext{2}{Energy Resolution at 2 keV}
\end{deluxetable*}
Lastly, the third objective of the SEUSHI instrument involves generating early alerts for coronal mass ejections. This is done by utilizing the EUV spectra, which exhibits coronal dimming, or reduction in emissions when plasma is rapidly evacuated into an erupting CME. Previous work has shown that quantitative measurements of coronal dimming (area, depth, slope of intensity drop) can be used to predict Earth-directed CME mass, speed, and potential geomagnetic impact before the CME reaches Earth (for example, \cite{mason_relationship_2016, mason_sdoeve_2019} and references therein). The EUV spectrograph of the SEUSHI instrument provides high resolution EUV spectra to measure coronal dimming, whose depth is proxy for CME mass and whose slope is proxy for CME velocity (e.g., \cite{mason_relationship_2016}). Additionally, multiple emission lines across ionization states of Fe in the spectral range of 17-22 nm, allow probing the temperature of the coronal plasma. Additionally, there is current on-going work to estimate density depletion using the density-sensitive line pair of Fe XIII at approximately 20 nm, using data from SDO-EVE. Further work is also ongoing, to explore what other parameters may affect the relationship of dimming with CMEs (e.g., the total unsigned flux, the amount of open field nearby, etc). Overall, the EUV spectra in this wavelength range provides a fertile ground for basic science to explore the relationship of the coronal plasma and its emission with the kinematics of the CME. The SEUSHI instruments spectral range covers the key part of the spectrum EUV spectrum that previous coronal dimming studies with SDO-EVE MEGS-A focused on, and with similar spectral resolution. The SEUSHI solar SXR images also provide flare location information that contributes to knowing the CME direction. The combination of these SEUSHI measurements can then provide alerts for earth directed CMEs which are missed by traditional white-light coronagraph instruments. The measurement requirements and corresponding instrument projected performance pertaining to the three objectives are listed in the last two columns of the STM. The SEUSHI instrument meets all measurement requirements for the science goals, other than the SNR requirement for coronal dimming (current SNR at 193 Å is approximately 200). One possible way to achieve better SNR is to combine adjacent Fe line of similar coronal temperature. For example, the coronal dimming iron lines include Fe IX 17.11 nm, Fe X 17.46 nm, Fe XI 17.72 nm, etc. which can be combined together to give better SNR at lower spectral resolution. Thus, the current value is sufficient for a the sounding rocket proof-of-concept test flight, and Section \ref{subsec:euv_concept} discusses multiple strategies for improving the SNR for future satellite version of the SEUSHI instrument. 

\subsection{SEUSHI Instrument Novelties and Sounding Rovcket mission goals}
\textbf{SEUSHI is designed as a low-power, low-mass, and compact instrument capable of generating flare alerts by leveraging the HOPE phenomenon. Its small form factor distinguishes it from larger instruments such as Hinode XRT and Yohkoh SXT, and makes it well suited for deployment on a range of platforms, including CubeSats and as an alert system for future lunar and Mars missions. A key novelty of SEUSHI is to perform onboard data processing, enabling real-time flare alerts based on the HOPE effect, as well as CME alerts using coronal dimming signatures. This allows alerts to be generated directly by the instrument without reliance on ground-based analysis.}

\textbf{For EUV coronal dimming, SEUSHI aims to achieve measurements comparable to those of SDO EVE MEGS-A, but with significantly reduced size, mass, and power requirements. Such capability would make it feasible to deploy on low-cost small satellites dedicated to space weather monitoring. Additionally, given the loss of MEGS-A due to CCD electronics failure, an instrument covering a similar wavelength range would restore access to critical spectral lines essential for solar observations.}

\textbf{Since the 10-minute SDO-EVE calibration sounding rocket flight is unlikely to observe a flare, the primary objective of this flight is to demonstrate SEUSHI measurement capability across three observing modes: pixel-level temperature and emission measure derivation, active-region soft X-ray spectroscopy, and EUV spectral irradiance observations. In the absence of a flare, validation will be achieved through quiescent-Sun and active-region observations that (a) validate pixel-level temperature and emission measure retrievals from SXR imaging, (b) demonstrate active-region spectral reconstruction through individual photon counting, and (c) verify expected signal levels for the EUV spectrograph.}

\section{Instrument Overview} \label{sec:instoverview}
\begin{figure*}[t!]
\centering
\includegraphics[width = \textwidth]{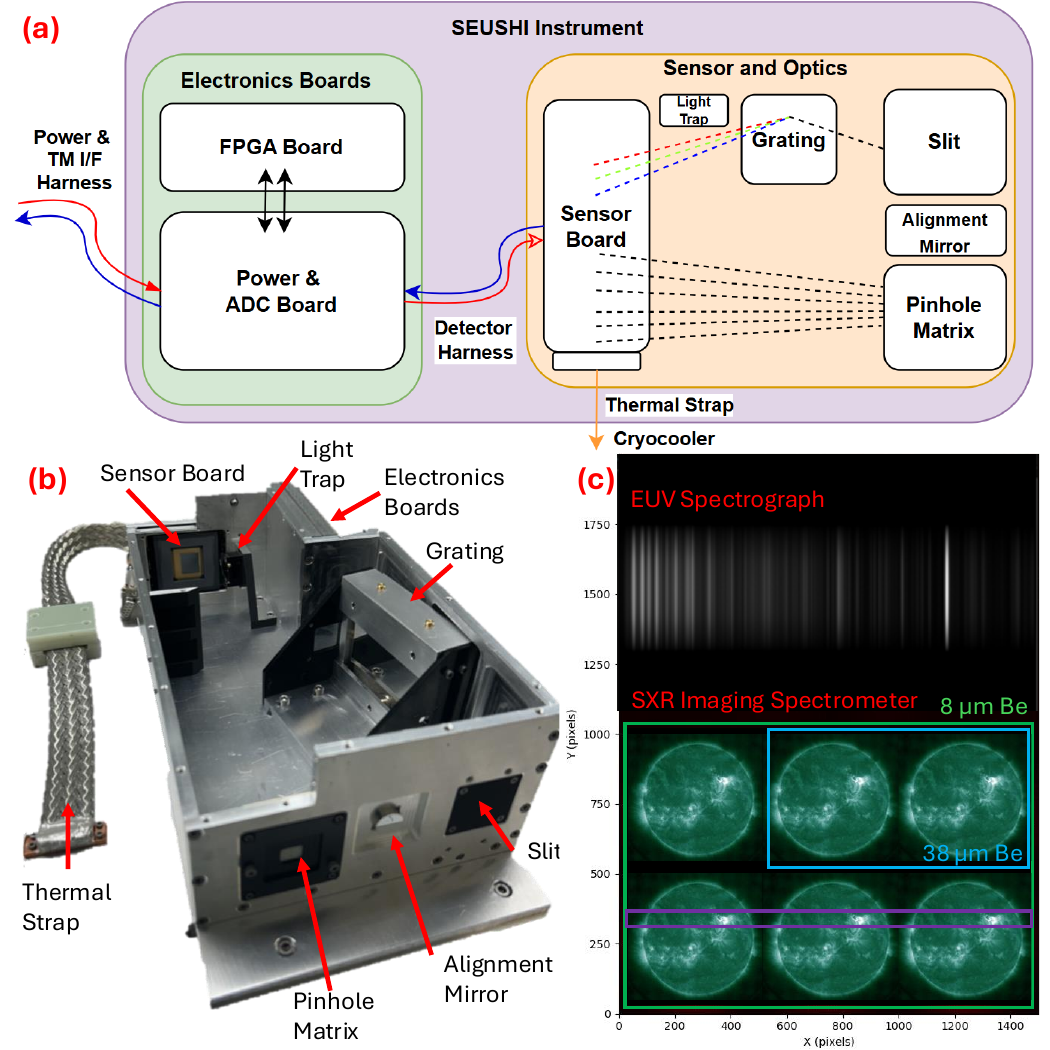}
\caption{(a) SEUSHI top-level block diagram showing different components of both the SXR imaging spectrometer and EUV spectrograph. The sensor and optics, including the slit, pinhole apertures and diffraction grating are shown in an orange box. The electronics unit including the FPGA board and the Power \& ADC board are shown in a green box. (b) SEUSHI flight model instrument without its top cover and different components labeled. (c) Shows a representational CMOS sensor image with the EUV spectrograph on the top and 6 pinhole images on the bottom. The purple box indicates 16 rows that will be readout at a fast rate of 100 Hz to enable photon-counting for determining the energy of incident X-ray photons.}
\label{fig:fig_1_overview}
\end{figure*}

%SXR Imaging Spectrometer Concept: Fig 1: SEUSHI Layout labelled diagram (pic with lables/ optics diagram)
Figure \ref{fig:fig_1_overview}(a) shows a block diagram representation of the SEUSHI instrument. The instrument consists of two parts which are: (a) a SXR imaging spectrometer called Solar Pinhole Imager of Coronal Irradiance with TUNAble-readout (SPICI-TUNA), and (b) an EUV spectrograph called TEMAKI, or The Euv Megs-A rocKet Instrument.  
The SXR imaging spectrometer consist of a pinhole matrix (with six pinholes in a 2 by 3 arrangement) along with thin-foil beryllium filters for generating full disk SXR images. Dotted black lines in the block diagram depict the photons from the pinholes hitting the sensor which is a back-illuminated complementary metal oxide semiconductor (CMOS) image sensor. The EUV spectrograph consists of a slit along with a thin-foil aluminum filter, with a diffraction grating for measuring the EUV spectra, and a zeroth order light trap, as shown in the orange box on the right side of the diagram. The front of the instrument also consists of an alignment mirror, that is used to coalign the SEUSHI instrument with the rocket reference optical alignment cube. The CMOS image sensor is readout by supporting electronics that includes two boards (i) an FPGA board and (ii) a Power \& ADC board. The digitized data from the instrument is then transferred to the rocket interface electronics using a synchronous-serial over Low Voltage Differential Signal (LVDS) interface, to be downlinked as reduced-resolution images. The instrument runs using a single 8-14 V power supply, with a power consumption of approximately 4 W. The FPGA board also consists of flash memory to store the full-resolution images onboard. In order to keep the sensor cool (target of -10$^\circ\mathrm{C}$ for minimizing thermal noise), the instrument also consists of a thermal strap that provides a thermal conduction path from the sensor to the cryo-cooler of the SDO-EVE sounding rocket.

Figure \ref{fig:fig_1_overview}(b) shows the flight model (without the top lid) of SEUSHI instrument with the various components labeled. The instrument chassis is machined using aluminium-6061 having a total mass of 4 kg, and dimensions of 350 mm x 165 mm x 110 mm. The key system level performance parameters of the SEUSHI instrument are summarized in Table \ref{tab:seushi_performance}. Figure \ref{fig:fig_1_overview}(c) shows a representational CMOS sensor image with the EUV spectrograph on the top and the 6 pinhole images on the bottom. The pinhole images have two sets of Be filters, with all 6 images having a 8 $\mathrm{\mu}$m Be filter (denoted by green box) and 2 images having a total of 38 $\mathrm{\mu}$m Be filter (denoted by blue box) that is realized using the combination of 8 $\mathrm{\mu}$m and 30 $\mathrm{\mu}$m Be filters. Additionally, a purple box indicates 16 rows that will be readout at a fast rate of 100 Hz, to enable photon-counting to determine the energy of incident X-ray photons.  The various components of the instrument, i.e., the SXR Imaging Spectrometer, EUV spectrograph, and sensor with electronics are described in the subsequent subsections. 
\begin{deluxetable*}{ll}[t!]
\tablecaption{SEUSHI Instrument Design Parameters \label{tab:seushi_performance}}
\tablewidth{0pt}
\tablehead{
\colhead{Parameter} & \colhead{Value}
}
\startdata
\multicolumn{2}{c}{\textbf{System-Level Parameters}} \\
\hline
Mass & 3.5 kg \\
Power & $3.7$ W \\
Volume & $350 \times 165 \times 110$ mm$^{3}$ \\
Supply Voltage & 8--14 V \\
Data Interface & Synchronous Serial over LVDS \\
Mechanical Interfaces & 4 $\times$ 8--32 bolts \\
Raw Data Rate (stored on flash) & 48.05 Mbits s$^{-1}$\\
% Rocket Telemetry Data Rate & 1.46 Mbits s$^{-1}$ (1000 x 752 binned images every 10 sec) \\
\hline
\multicolumn{2}{c}{\textbf{Soft X-ray (SXR) Imaging Spectrometer}} \\
\hline
% Energy Range & 1.1--6.8 keV \\
% Field of View & $1.3\,R_{\odot}$ \\
% Spatial Resolution & $57''$ \\
% Energy Resolution & 0.08 keV (Photon-counting mode) \\
% Spectral Bandpasses & 2 (For generating Temp. and E.M. maps)\\
Filter Material & Beryllium\\
Filter Thickness & 8 $\mathrm{\mu}$m, 38 $\mathrm{\mu}$m\\
Pinhole Aperture Plate Material & Tungsten\\
Pinhole to CMOS Sensor Distance & 274 mm \\
Pinhole Aperture Diameter & 75 $\mathrm{\mu}m$\\
Integration Time (Full Images) & 5 seconds\\
Integration Time (Fast Readout) & 10 milliseconds \\
% Signal to Noise Ratio & 38.9 at 2 keV \\
% Accuracy & $\mathrm{\approx}$10\% \\
\hline
\multicolumn{2}{c}{\textbf{EUV Spectrograph}} \\
\hline
% Bandpass & 16.1--33.8 nm \\
% Spectral Resolution & $\leq 0.18$ nm \\
% sensor Format & Spectral: 1.5k pixels (1D) \\
% Signal-to-Noise Ratio & $>400$ for bright lines \\
% Dynamic Range & Quiet Sun to active Sun \\
Slit to Grating Distance & 106.45 mm \\
Grating to Sensor distance & 190.28mm \\
Slit Dimensions & 0.025 mm x 1.0 mm \\
Filter Material \& Thickness & C/Al/C 50/2016/50 $\mathrm{\AA}$ \\
Slit Aperture Plate Material & Stainless Steel\\
Angle of Incidence at Grating & 78.19$\mathrm{^\circ}$\\
Angle of Diffraction & 73.65$\mathrm{^\circ}$\\
Angle of Incidence at Detector & 28.16$\mathrm{^\circ}$\\
Integration Time & 5 s \\
\enddata
\tablecomments{The raw data rate is calculated considering both the slow and fast readout of the pixels. This includes 16 x 1504 pixels that are readout every 10 milliseconds and 1984 x 1504 pixels that are readout out every 5 seconds. Each pixel value is stored as a 16 bit integer.}
\end{deluxetable*}

\subsection{SXR Imaging Spectrometer} 
\begin{figure}[t!]
    \centering
    \includegraphics[width=\linewidth]{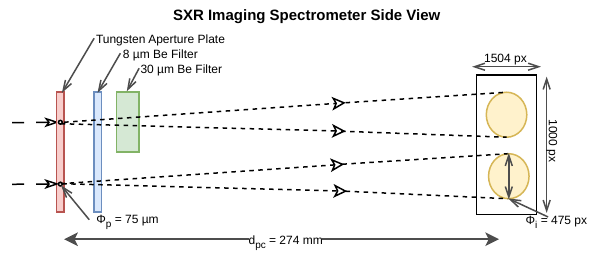}
    \caption{Diagram depicting the SXR imaging spectrometer with only two of the six pinholes shown simplicity. \textbf{The dashed line with arrows indicate the direction of X-rays entering the instrument.} The pinhole aperture tungsten plate is shown in the left (in red), 8 micron beryllium filter (blue), and 30 micron beryllium filter (green) are shown in the figure. The pinhole images are shown on the right side in yellow, which are formed on the sensor. The images occupy an area of 1000 x 1504 pixels on the detector and the diameter of each image is approximately 475 pixels.}
    \label{fig:seushioptics_sxr}
\end{figure}

Figure \ref{fig:seushioptics_sxr} depicts the SXR imaging spectrometer, that consists of a pinhole matrix with 6 pinholes made in a tungsten sheet. Tungsten is used as it is effective in blocking Hard X-rays during large flares. The pinhole aperture consists of a larger diameter aperture made in an aluminum sheet, and a smaller diameter aperture made in a tungsten sheet. The tungsten aperture sheet (which is 50 $\mathrm{\mu}$m thick)  was ordered from Thorlabs, with 6 pinholes arranged in a 2x3 grid. The diameter of the pinhole aperture is 75 $\mathrm{\mu}$m which is chosen to ensure enough signal to noise, for the limited duration of the sounding rocket flight (discussed further in Appendix \ref{sec:appendixA}). Similarly, the tungsten aperture plate thickness is chosen for the rocket flight active regions, a higher tungsten plate thickness will be used for flare observations to suppress the Hard X-ray signal, for a future satellite-version of the SEUSHI instrument. Behind the apertures there are beryllium filters of different thicknesses, that are used to create different bandpasses within the SXR range. The current design uses two thicknesses which are 8 micron and 38 micron (which is created by overlapping an 8 micron and 30 micron foil filter), and this is shown pictorially in Figure \ref{fig:seushioptics_sxr}. The distance between the pinhole apertures and the sensor is 274 mm, which leads to a solar image spanning 200 pixels on the sensor. The design optimization of the pinhole distance, filter thicknesses, and aperture size are discussed in Section \ref{subsec:sxr_concept}. 
\subsection{EUV spectrograph}
\begin{figure}[t!]
    \centering
    \includegraphics[width=\linewidth]{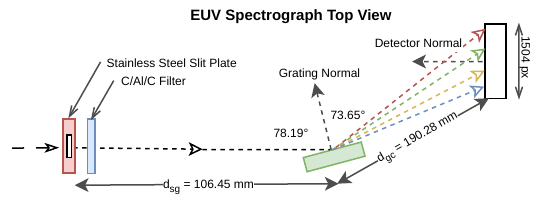}
    \caption{Diagram depicting the EUV spectrograph. \textbf{The dashed line with arrows indicate the direction of EUV light entering the instrument.} The slit is on the left in a stainless steel plate and this is followed by a C/Al/C filter. After this the light hits the diffraction grating at grazing incidence, and then reaches the CMOS image sensor forming the EUV spectrograph image. The EUV spectrograph also consists of a zeroth-order light trip (not shown in the picture) to prevent zeroth-order from the grating reaching the CMOS sensor.}
    \label{fig:seushioptics_euv}
\end{figure}

Figure \ref{fig:seushioptics_euv} depicts the EUV spectrograph design. The spectrograph operates at grazing incidence and is similar to the design of the SDO-EVE MEGS-A channel \citep{woods_extreme_2012}. Detailed optical design of the instrument is described in \citep{crotser_sdo-eve_2007}, and only key details are summarized here. The EUV spectrograph optics consist of a slit in a stainless steel sheet, followed by a thin-foil C/Al/C filter (with 50/2016/50 angstrom thickness). The foil filter blocks visible light while transmitting EUV light. The light then reaches the diffraction grating at grazing incidence, with an angle of 78.19 degrees. The diffracted light (with a central angle of diffraction of 73.65 degrees) from the grating hits the top half of the same CMOS image sensor as the SXR pinhole images, forming the spectrograph image. A zeroth-order light trap is also added to the instrument, to prevent the zeroth order light from reaching the sensor and to reduce visible scatter in the instrument. Signal estimates and design details of the EUV spectrograph are discussed in more detail in Section \ref{subsec:euv_concept}.    
\subsection{CMOS Image Sensor}
The SEUSHI instrument employs the Compact Spectral Imaging Electronics (CSIE), a modular sensor and electronics system developed at the Laboratory for Atmospheric and Space Physics for use in multiple small-satellites (including Sun Coronal Ejection Tracker (SunCET) and the CubeSat Imaging X-ray Solar Spectrometer (CubIXSS) missions). SEUSHI utilizes the Teledyne e2v CIS115 back-illuminated CMOS image sensor, which offers low noise (mean electron noise of 5 electrons rms per pixel) and fast readout (typical rates of 6.2 mega-pixels per second) capabilities suitable for both imaging and photon-counting spectroscopy. A CMOS image sensor is better suited to this type of instrument than a charged-coupled device (CCD) because of the fast readout speed, low power dissipation and ability to arbitrarily readout individual rows at a high rate. The CIS115 sensor is a back-thinned device, enabling it to operate in both soft X-ray (SXR) and extreme ultraviolet (EUV) wavelengths. It is also radiation-hardened and is currently flying aboard the JANUS instrument on the JUICE mission to Jupiter \citep{palumbo_janus_2025}. The CIS115 sensor has dimensions of 1504 x 2000 pixels, with a pixel size of 7 x 7 um$^2$. The sensor Silicon thickness is 30 microns with a Silicon Oxide layer of 0.07 microns. The CSIE system consists of three primary boards: a sensor board, a field-programmable gate array (FPGA) board, and a combined power and analog-to-digital conversion (ADC) board. A previous generation of CSIE electronics have sounding rocket flight heritage onboard the Compact SOLar STellar Irradiance Comparison Experiment (SOLSTICE or CSOL) and Polar-NOx missions \citep{thiemann_solar_2023, bailey_sounding_2022}. The FPGA and power/ADC boards together form the electronics stack, which accounts for the majority of the instrument’s power dissipation. To operate the sensor at a low temperature (-10$^\circ\mathrm{C}$) a thermal strap is designed that provides a thermal conduction path between the sensor and the sounding rocket's cryocooler that operates down to -120$^\circ\mathrm{C}$. Data are transmitted via a synchronous serial interface to the rocket telemetry system through the electrical ground support equipment. The total power dissipation of the electronics is approximately 3.8 W, and includes both the CMOS sensor and supporting electronics. 

\section{Instrument Design} \label{sec:design}
This section describes the design of both the SXR imaging spectrometer and the EUV spectrograph including detailed discussion of their different design parameters and how those effect the measurement capabilities. Signal models of both the SXR imaging spectrometer (using CHIANTI \citep{del_zanna_chiantiatomic_2021} model spectra) and EUV spectrograph (using previous SDO-EVE spectra) are also discussed. 
\subsection{Soft X-ray Imaging Spectrometer} \label{subsec:sxr_concept}
The main instrument performance parameters of the SXR imaging spectrometer are the spatial resolution, spectral resolution, and the energy range. These depend on the design parameters which include pinhole aperture diameter, sensor pixel size, pinhole to sensor distance, filter material and thickness. The sensor readout rate and sensor ADC gain are also important design parameters for detecting the energy of individual SXR photons and for energy resolution, respectively. Appendix \ref{sec:appendixA} describes the relation between the instrument design parameters and output performance. The main constraint for the sounding rocket version of the SEUSHI instrument was the pinhole to sensor distance, which is limited by the space available on the sounding rocket. A pinhole to sensor distance of 274 mm was chosen which leads to a image diameter of 475 pixels (considering 1.3 times the radius of the sun). In this configuration, 3 solar images can be stacked side-by-side with different passbands. Additionally, for the SEUSHI instrument, one row of three images are created to allow photon counting in one row and filter-ratio method in the other row. The pinhole-sensor distance of 274 mm provides pixel scale of 5.3 arcseconds, and pinhole diameter of 75 $\mathrm{\mu}$m provides spatial resolution of 57 arcseconds (11 pixels).  While this spatial resolution can resolve active regions and flares, it cannot resolve fine coronal loop features. Such a design is sufficient for the technology demonstration flight to retrieve HOPE signatures using the ratio method as well photon counting to obtain SXR spectra. 
\begin{figure}[!htbp]
    \centering
    \includegraphics[width=\linewidth]{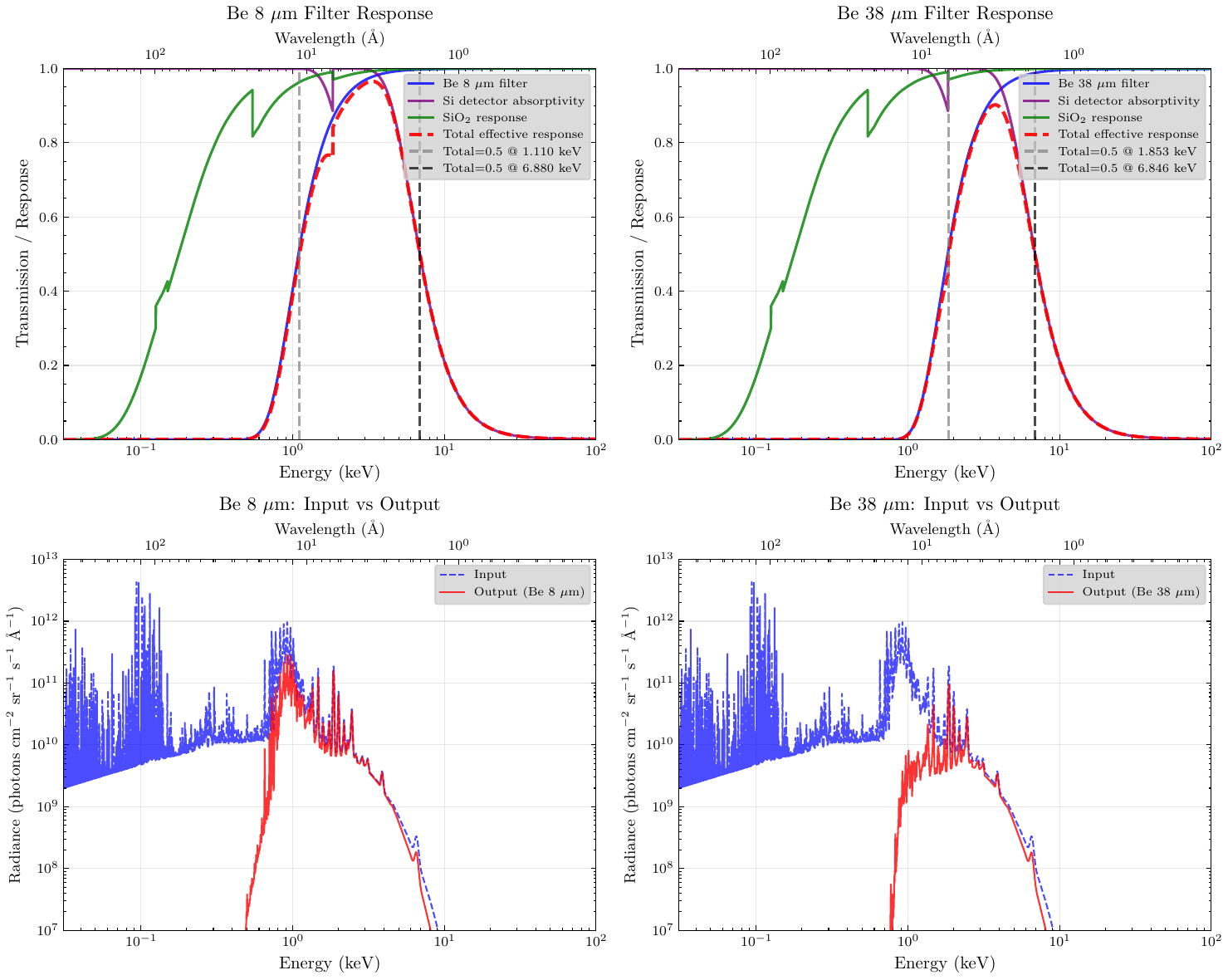}
    \caption{SXR imaging spectrometer filter response and signal estimates. The top row shows the Be filter transmittance for the 8 micron and the 38 micron filters in the left and right panels respectively (denoted by blue solid lines). The absorptance of Si sensor (purple line), Silicon Oxide transmittance (green line), and total effective response (dashed red line) is also shown. The dashed gray and black vertical lines indicate the lower and upper energies where the transmission is 0.5, indicating the bandpass of the two channels. The bottom row shows the input solar spectrum for a 10 MK isothermal CHIANTI spectrum (dashed blue line), and output spectra considering the response of the Be filter, silicon sensor absorptance, and silicon oxide response (solid red line) for both the 8 micron and 38 micron Be filters.}
    \label{fig:seushi_sxr_filter}
\end{figure}
A future version of the instrument, for example on a satellite platform, can utilize larger pinhole distance to further improve spatial resolution. The spectral range of the SXR imaging spectrometer is decided by the material and thickness of the thin-foil filters placed after the pinhole apertures. 

Figure \ref{fig:seushi_sxr_filter} shows the filter response of the instrument. In order to model the filter response, a model isothermal solar spectra is generated using the CHIANTI \citep{del_zanna_chiantiatomic_2021} atomic database. The response is then convolved with the filter response that is obtained from XrayDB. The sensor quantum efficiency and gain is then accounted for to get the number of electrons generated as signal within the sensor. Due to constraints in purchasing beryllium as well as schedule considerations, existing spare thin-foil filters from previous GOES-XRS instruments built at LASP were chosen. These include an 8 micron filter for four of the six channels, and 8 micron together with 30 microns (resulting in 38 microns) for remaining two of the six channels. The response of these filters in the top left and top right panels of Figure \ref{fig:seushi_sxr_filter} for the 8 micron and 38 micron filters respectively. The effective response including the filter transmittance, the sensor silicon absorptance and Silicon oxide layer transmittance are shown in the top row plots. A model input spectrum is generated from CHIANTI for a 10 MK isothermal plasma source, and is shown in the bottom panels in blue. The measured signal considering the effective filter and sensor responses are shown in the bottom row for the 8 micron and the 38 micron beryllium filters. 

\begin{figure}[!t]
    \centering
    \includegraphics[width=\linewidth]{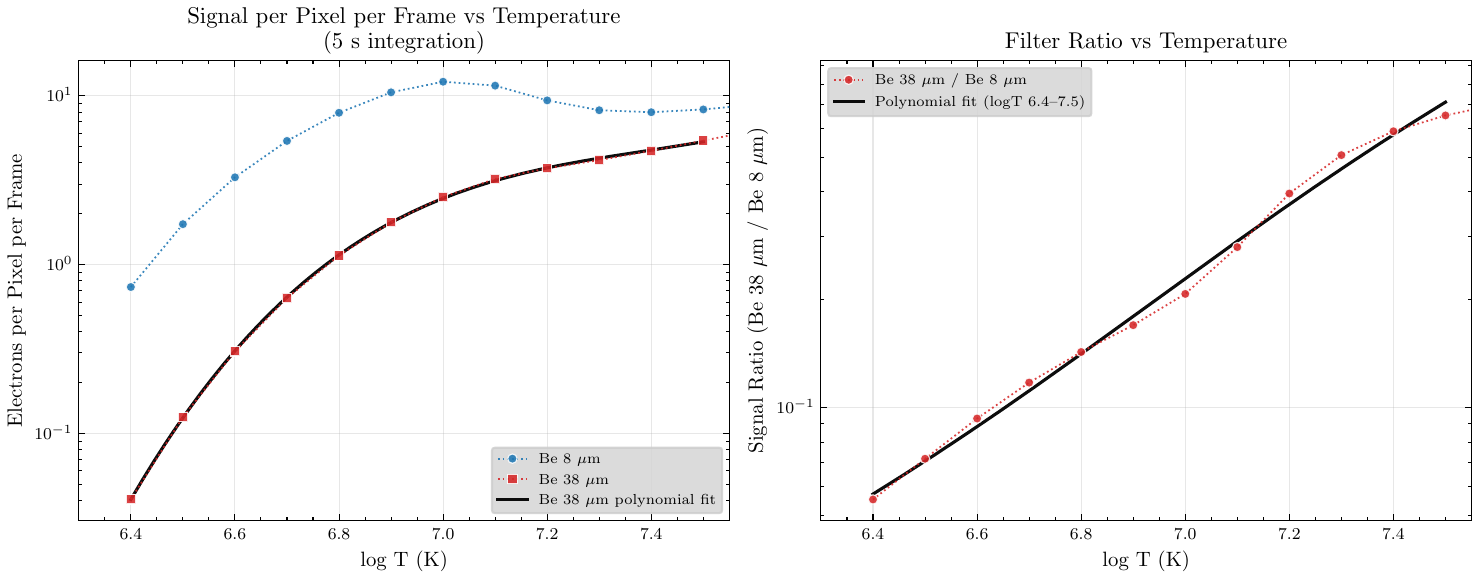}
    \caption{Left panel shows the total signal (in electrons per second) through both the 8 micron (blue) and 38 micron (red) beryllium filters, as a function of solar plasma temperature from logT=6.4 (2.5 MK) to logT=7.5 (31 MK). The right panel shows the filter ratio, as a function of the plasma temperature. The curve is also fitted to a polynomial (dashed black line) which can be used to compute plasma temperature from measured ratio.}
    \label{fig:seushi_temperature}
\end{figure}

One application of the SEUSHI SXR images is to obtain temperature and emission measure map images. This is similar to the GOES XRS-A/XRS-B ratio method but applied to the imager pixel resolution. For the SEUSHI SXR imager, the ratio of the 38 micron to 8 micron channels can be used to create temperature and emission measure maps. In order to do this, different temperature isothermal plasma spectra were generated using CHIANTI over a range of \textbf{logT = 6.4 ($\approx2.5$ MK) to logT = 7.4 ($\approx31$ MK)} (where T denotes temperature in Kelvin), with a fixed emission measure of $\mathrm{10^{27}cm^{-3}}$. Then, for each spectrum the ratio of the irradiance measured was computed, considering the response of the filter, silicon sensor absorptance, as well as the pinhole aperture area and sensor gain to obtain the signal in electrons. Since the CHIANTI isothermal spectra are at fixed arbitrarily chosen emission measure ($\mathrm{10^{27}cm^{-3}}$), the electron signal estimates are not accurate and only used to calculate the ratio. Figure \ref{fig:seushi_temperature} left panel shows the signal in electrons (for an integration time of 5 second) measured through both the 8 micron and 38 micron Be filter. A polynomial fit to the 38 micron filter signal is also shown using a dashed black line. The right side panel of the plot shows the computed ratio and a polynomial fit to the same.  
It can be observed that the polynomial is monotonically increasing, which implies that a measured ratio can be inverted to obtain plasma temperature by using the polynomial fit as follows: 

\begin{equation}
T_{log} = 0.192186 \cdot R^{3} + 0.406147 \cdot R^{2} + 1.233478 \cdot R + 7.673471
\end{equation}

where $T_{log} = \log_{10}(T(K))$ (valid for $6.4 \leq T_{log} \leq 7.5$), $R = \log_{10}(\frac{I_{38\mu m}}{I_{8\mu m}})$, $I_{38\mu m}$ denotes the measured signal (in electrons per pixel per frame) in the 38 micron channel and $I_{8\mu m}$ denotes the measured signal in the 8 micron channel. To determine the emission measure, the Be 38 $\mu$m filter signal can first be determined from the temperature using the following polynomial fit:
\begin{equation}
\log_{10}(I_{38\mu m, measured}) = 1.645121 \cdot T_{log}^{3} -36.475049 \cdot T_{log}^{2} + 270.040990 \cdot T_{log} -666.893703
\end{equation}

Then, a ratio of the measured signal to the modeled signal can be used to determine the emission measure per pixel.
\begin{equation}
    EM(cm^{-3}) = \frac{I_{38\mu m, measured}}{I_{38\mu m, model}}
\end{equation}
Thus, when the instrument is launched, the measured ratio can be used together with this polynomial to compute the spatially resolved plasma temperature and emission measure map. This methodology has been used extensively using full-disk GOES-XRS data (for example see \cite{telikicherla_improving_2025} and reference therein), but with SEUSHI this can be extended to obtain plasma parameters at the imager resolution.

\subsection{EUV Spectrograph Design Optimization} \label{subsec:euv_concept}
\begin{figure}[!t]
    \centering    \includegraphics[width=\linewidth]{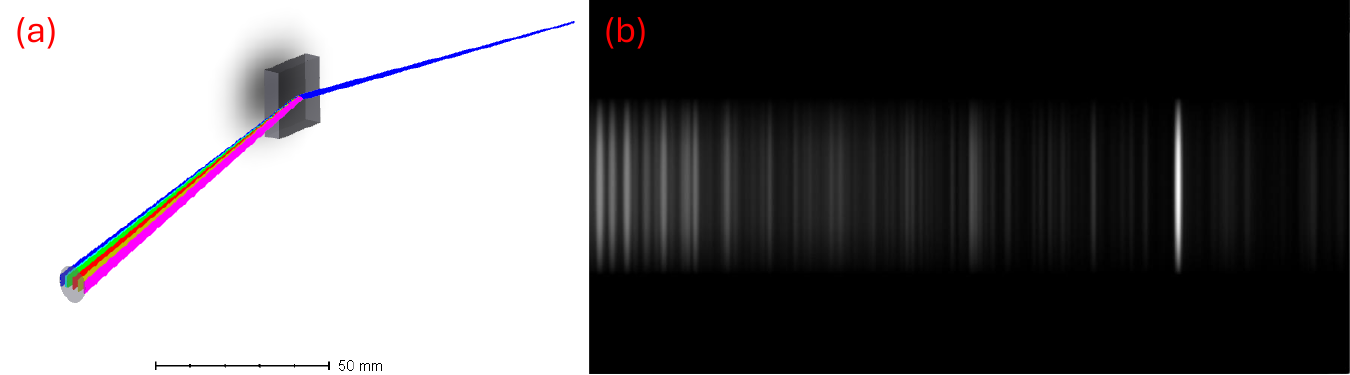}
    \caption{(a) The ray trace model of the grating created using Zemax, (b) Simulated sensor spectrograph image generating using Zemax corresponding to a slit size of 0.025 mm x 1.0 mm. The signal spans approximately 500 rows in the vertical dimension, and the entire sensor (1504 pixels) in the horizontal dimension. The solar spectra is generated using CHIANTI and then passed through the instrument model considering slit geometry, grating efficiency, filter transmission, and sensor quantum efficiency.}
    \label{fig:euvdesign}
\end{figure}

\begin{figure}[!htbp]
    \centering 
    \includegraphics[width=\linewidth]{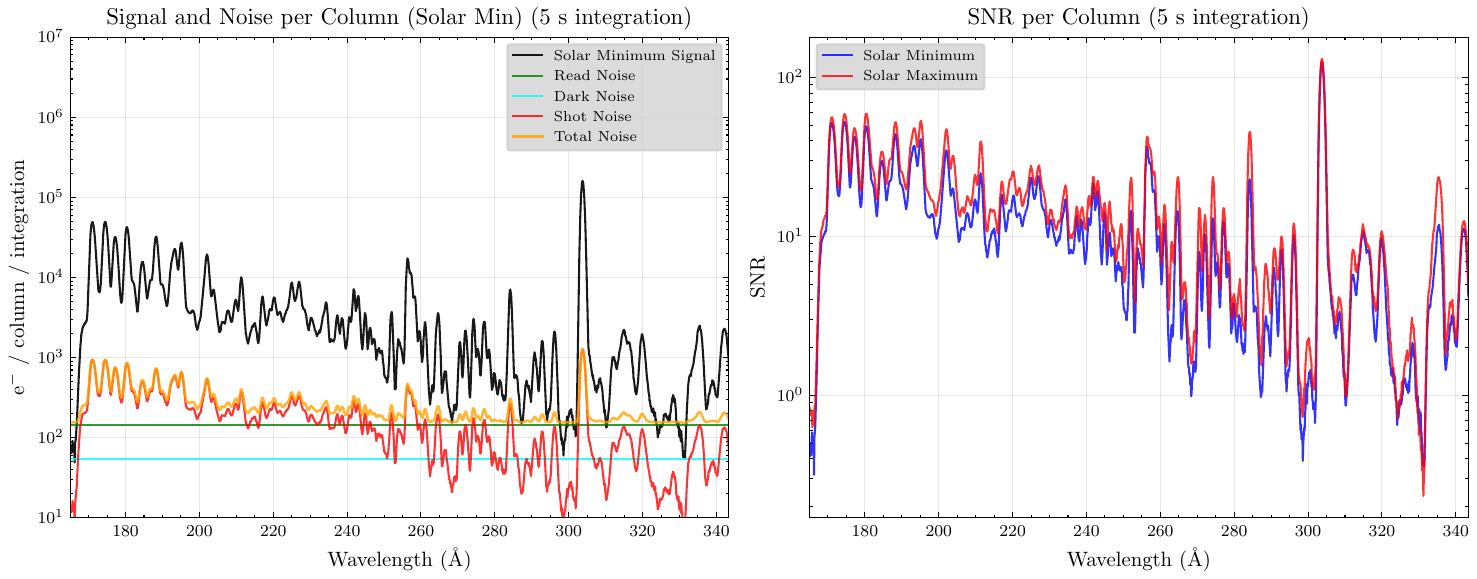}
    \caption{Signal and noise estimates using model spectra generated with CHIANTI. Left plot shows the signal (black), read noise (green), dark noise (blue), shot noise (red), and total noise (yellow) for summed over all rows of a particular spectrograph column. estimates for solar minimum spectra. Right plot shows the Signal to Noise Ratio (SNR) for both solar minimum (blue) and solar maximum (red) spectra.}
    \label{fig:euv_signal}
\end{figure}

The main instrument performance parameters of the EUV spectrograph are the spectral range and spectral resolution. These depend on the grating parameters, thin-foil filter material and thickness, as well as the grating-sensor geometry. The design optimization is performed by creating a ray-trace model of the spectrograph using Zemax (also referred to as Ansys OpticStudio \cite{OpticStudio2025}), which is shown in Figure \ref{fig:euvdesign}(a). The ray trace is used to optimize the geometry of the spectrograph, optimizing spectral focus, trading spatial focus quality for improved spectral resolution. This includes optimizing the distance between the slit and grating as well as between the grating and the sensor, based on the grating properties. Figure \ref{fig:euvdesign}(b) shows a simulated sensor spectrograph image that was generated using the ray trace program by feeding in a solar spectra (shown in Figure \ref{fig:euv_signal}) convolved with the instrument response function. Signal estimates are generated using SDO-EVE solar minimum and solar maximum input spectra, and the corrected for slit geometry (considering a 0.025 mm x 1.0 mm rectangular slit), filter transmission, grating efficiency, detector responsivity, and integration time. The resulting image has a vertical spread of about 500 pixels and a horizontal spread spanning the entire detector, which is 1504 pixels.
Figure \ref{fig:euv_signal} shows the signal and noise estimates of the EUV spectrograph. The signal is summed over each column of the spectrograph and plotted as the total signal (black line) vs wavelength. This spans a wavelength range of 165 $\mathrm{\AA}$ to 343.4 $\mathrm{\AA}$. The different noise sources including shot noise (red line), dark noise (blue line), and read noise (green line) are shown in the left panel of Figure \ref{fig:euv_signal}. The total noise of an entire column (yellow line) is computed by the root mean squared sum of the total noise in each pixel of a column. A detailed discussion of the noise calculations are given in Appendix \ref{sec:appendixA}. 

The Signal to Noise Ratio (SNR, right panel of Figure \ref{fig:euv_signal}) is plotted for both solar minimum (blue line) and solar maximum (red line) SDO-EVE spectra. The SNR is approximately 32.7 per column at an example line of 193$\mathrm{\AA}$ for a 5 second integration time. To compute the SNR for the line feature, this value is multiplied by $\mathrm{\sqrt{W_{line}}}$, which denotes the line-width and is approximately 3 pixels. Additionally, the SNR can also be further improved by  co-adding images over a 60 second duration, which is a multiplication factor of $\mathrm{\sqrt{60/5}}$, and leads to a SNR of approximately 196. The science traceability matrix lists (Table \ref{tab:stm}) lists an SNR of about 400 for bright lines required for coronal dimming. This requirement arises since coronal dimming measurements require good precision of measurement to detect a 1-3\% change in signal (for example see, \citep{mason_relationship_2016, mason_sdoeve_2019}). A higher SNR is likely to have better precision since the error-bars around the measurement are lower. A future improved satellite version of the SEUSHI instrument could incorporate a slit that is twice as wide and twice as long. These increased slit dimensions would leading to a factor of two ($\mathrm{\sqrt{4}}$) improvement in signal-to-noise ratio, from approximately 200 to 400.

\section{Alignment and Calibration}
The alignment of the SEUSHI instrument was carried out to co-align the SXR pinhole apertures with the EUV spectrograph slit, ensuring that when the sounding rocket is pointed at the Sun, both instruments are simultaneously aligned to the solar disk. This section first details the alignment procedure for the SXR Imaging Spectrometer, followed by the alignment and wavelength calibration of the EUV Spectrograph. The planned calibration approach for the SXR Imaging Spectrometer is also outlined.
\label{sec:allign_cal}
\subsection{SXR Imaging Spectrometer Alignment}
\begin{figure}[!htpb]
    \centering
\includegraphics[width=\linewidth]{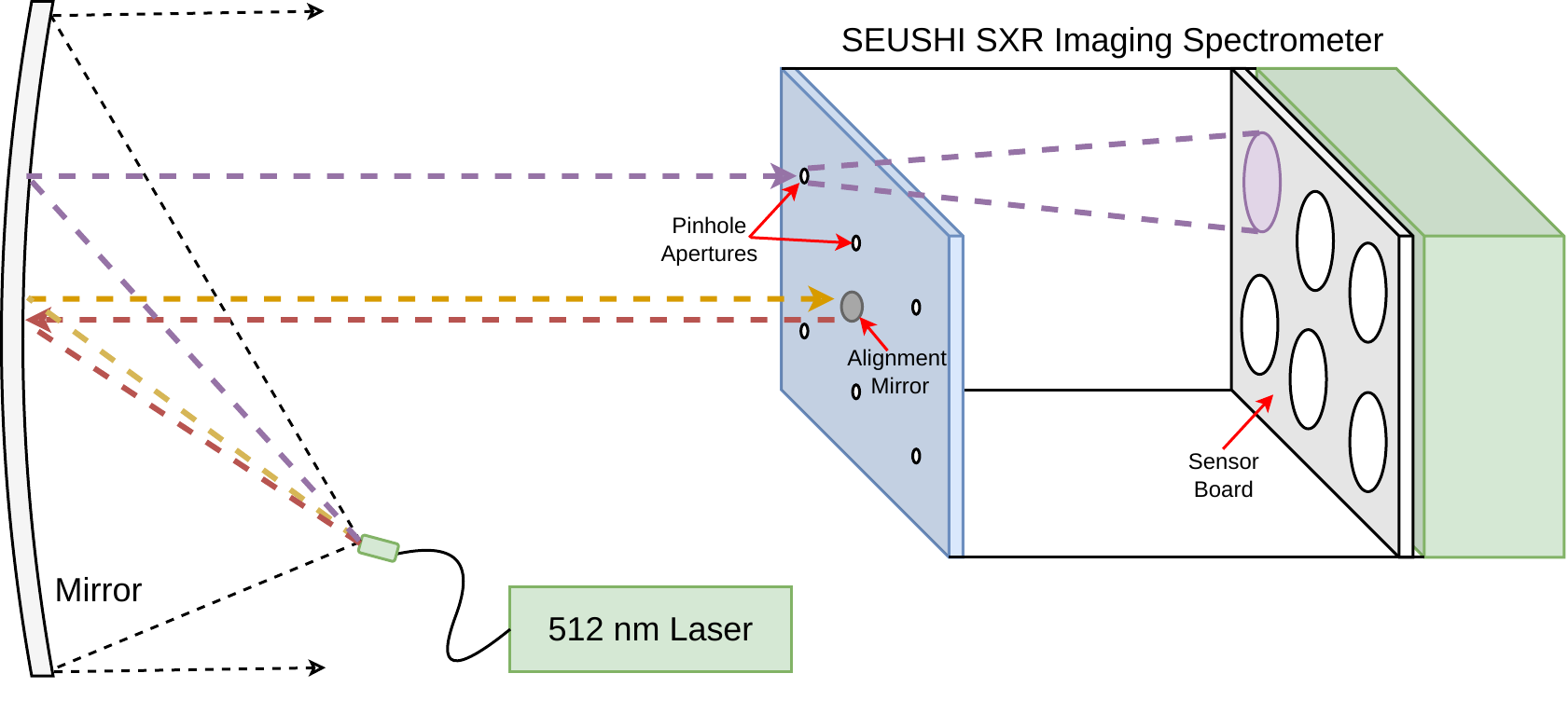}
    \caption{Sketch of SXR imaging spectrometer alignment setup with the 512 nm laser source, collimating mirror, and SEUSHI instrument. A retroreflection from the alignment mirror on the SEUSHI instrument was used to align the source with the instrument. An example pinhole image is shown in purple.}
    \label{fig:laser_images}
\end{figure}

\begin{figure}[!t]
    \centering
\includegraphics[width=\linewidth]{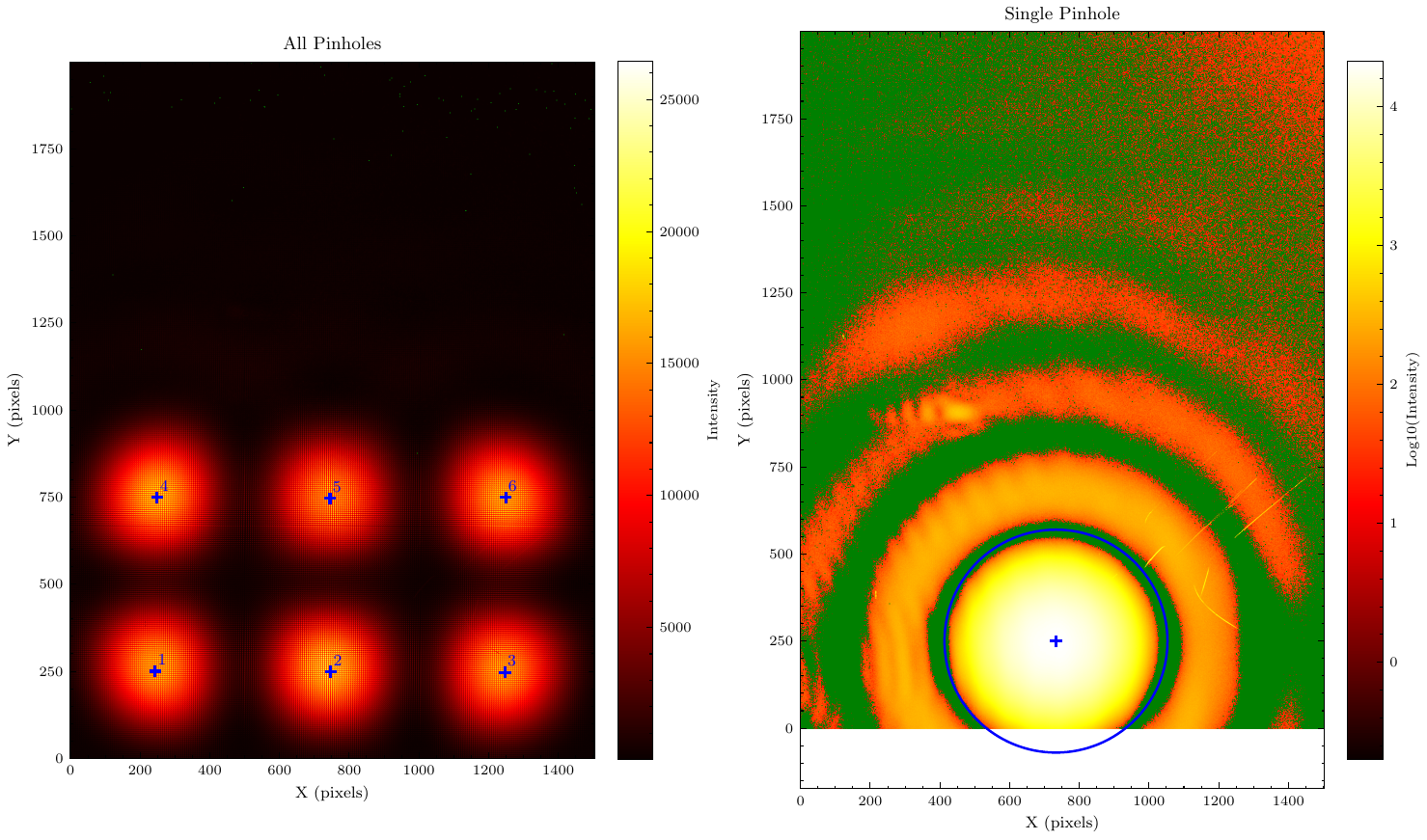}
    \caption{The left panel shows an alignment test image with all six pinholes exposed to the laser, plotted in linear scale. The centers of the 6 different pinhole images are also labeled 1-6. The right panel shows a test image with only one pinhole (\#2) exposed to the laser source, plotted in log scale. The first Airy disk is indicated using a blue circle. Dark images have been subtracted for both images, and green dots indicate where the signal after dark subtraction was 0 or negative.}
    \label{fig:sxr_pinhole_image}
\end{figure}

The alignment for the pinholes was performed using a 515 nm laser source after the Be filter was removed to let visible light reach the sensor. Figure \ref{fig:laser_images} shows the experimental setup in which light is directed to a large collimating mirror to generate a collimated beam wider than the instrument.  
This beam then reaches the pinhole apertures to create an image on the sensor. The SEUSHI instrument also consists of an alignment mirror that is used to position the laser source correctly: when aligned correctly an image of the source is created back at the laser as depicted in the figure. The test was performed in different configurations, in the first configuration the 8 $\mu$m Be and 38 $\mu$m Be filter were removed to allow 515 nm laser to reach the detector. In the second configuration only one pinhole was illuminated at a time by manually blocking all the other pinholes. However during this test, only the 8 $\mu$m Be was removed so that only 4 of the 6 pinholes could be illuminated in this way (due to the logistical constraints of removing the 30 micron filter, which is accessible from the interior of the instrument). The left panel of Figure \ref{fig:sxr_pinhole_image} shows an laser image taken during the alignment test, with six pinholes illuminated. Dark subtraction is applied to the image and green dots indicate where the signal after dark subtraction was 0 or negative. Then, the six pinhole centers (marked with blue labels from 1-6) were computed by fitting a circle to the images. Table \ref{tab:seushi_pinholes} lists the designed and measured center pixel location for each of the six images. The pinhole aperture sheet is designed as a 2x3 grid with grid spacing of 3500 $\mathrm{\mu}$m (500 pixels), with the bottom left image at (250, 250) pixels. The average offset center offset, i.e., design $-$ measured center coordinates (or ($\Delta x$, $\Delta y$)) is (3.33, 1.67) pixels. The right panel of Figure \ref{fig:sxr_pinhole_image} shows a test image taken with only one pinhole aperture (pinhole 2) illuminated. The diffraction pattern of the pinhole image is also visible as well as the Airy disk. 

\begin{deluxetable}{crrcc}[!htbp]
\tablecolumns{5}
\tablewidth{0pt}
\tablecaption{Laser Alignment Test Results\label{tab:seushi_pinholes}}
\tablehead{
\colhead{Pinhole \#} &
\colhead{\shortstack{Designed Center\\$(x, y)$}} &
\colhead{\shortstack{Measured Center\\$(x, y)$}} &
\colhead{\shortstack{Airy Disk Diam.\\(px)}} &
\colhead{\shortstack{Calculated Aperture\\($\mu$m)}}
}
\startdata
1 & (250, 250)   & (243.0, 250.0)   & 650.0 & 75.67 \\
2 & (750, 250)   & (747.0, 249.0)   & 640.0 & 76.85 \\
3 & (1250, 250)  & (1247.0, 247.0)  & 620.0 & 79.33 \\
4 & (250, 750)   & (249.0, 749.0)   & 676.0 & 72.76 \\
5 & (750, 750)   & (745.0, 746.0)   & \nodata & \nodata \\
6 & (1250, 750)  & (1249.0, 749.0)  & \nodata & \nodata \\
\enddata
\tablecomments{Average center offset (design $-$ measured): ($\Delta x$, $\Delta y$) = (3.33, 1.67) px. Average calculated aperture diameter: $76 \pm 2~\mu$m. The Airy disk of pinholes 4 and 5 were not measured as they were covered by the 30 $\mathrm{\mu}m$ Be filter.}
\end{deluxetable}

The first Airy disk radius was computed by finding the local minima of intensity closest to the center of the image. The observed Airy disk radius was used to compute the pinhole aperture diameter. Table \ref{tab:seushi_pinholes} lists the designed and measured pinhole aperture sizes for four pinholes tested in this configuration. The average measured aperture diameter is $76 \pm 2~\mu$m, which indicates that the designed 75$\mathrm{~\mu}$m value is within the error bars of the measurement.

\subsection{EUV Spectrograph Alignment and Wavelength Calibration}
The grating alignment was performed with the same laser light source as the pinhole alignment system, using the pinhole locations to establish the instrument boresight. Then, the grating was adjusted such that zeroth order was aligned with an alignment target. This ensured the boresight of the spectrograph was co-aligned with the pinhole imager.
\begin{figure}[!htpb]
    \centering
    \includegraphics[width=\linewidth]{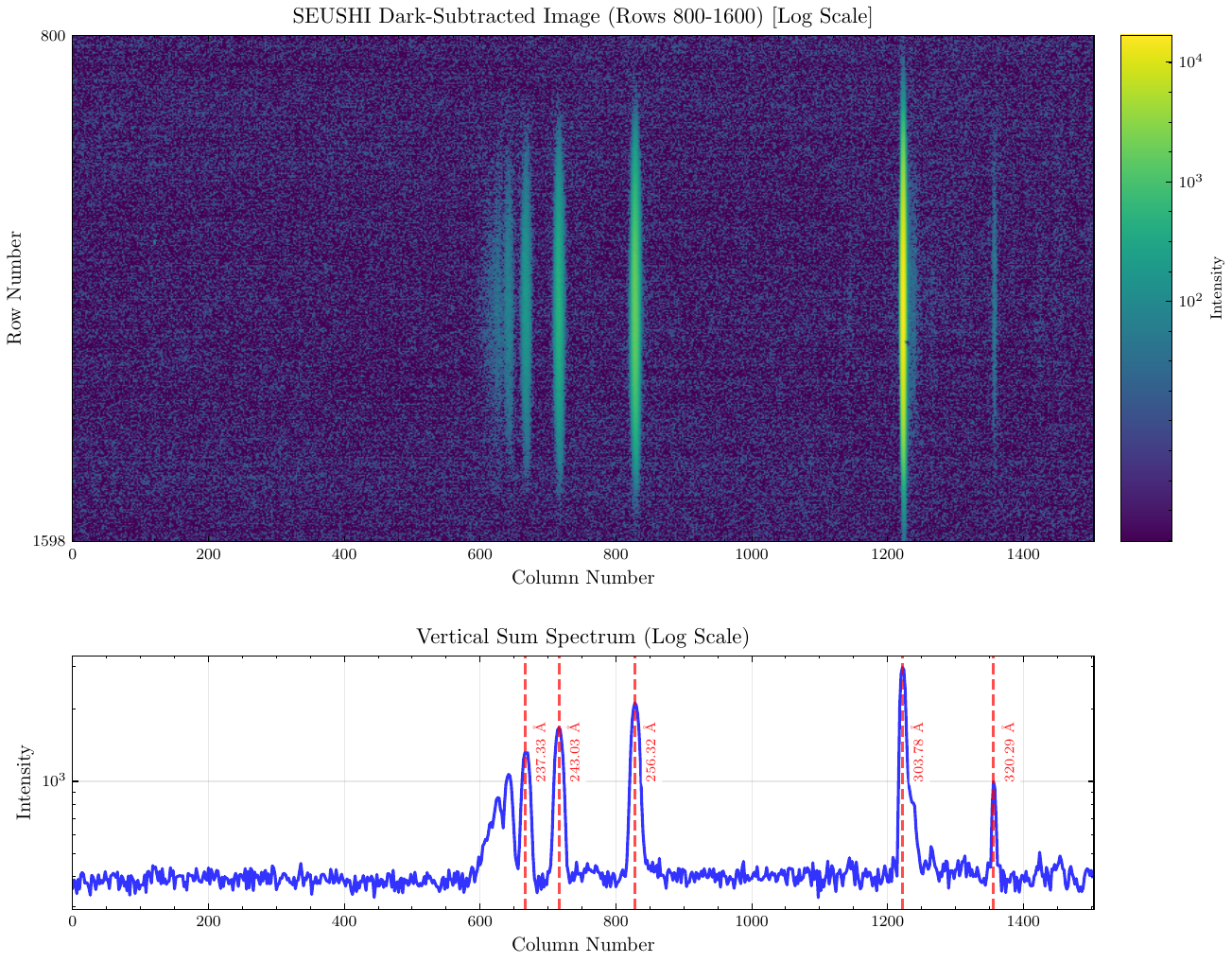}
    \caption{EUV spectrograph calibration measurements. The top panel shows the spectrograph image taken off a He-II hollow cathode EUV lamp. Dark subtraction is applied to the image, and it is normalized and plotted in log scale. The bottom panel shows the vertical sum spectrum showing the various He-II spectral lines.}
    \label{fig:euv_cal}
\end{figure}
A preliminary EUV spectrograph wavelength calibration was performed using an in-band He II hollow cathode lamp in vacuum to validate the visible light alignment test and to provide wavelength scale calibration. The top panel of Figure \ref{fig:euv_cal} shows the EUV spectrograph image taken during the test, which has been dark subtracted and plotted on a log scale. The bottom panel of the Figure \ref{fig:euv_cal} shows the intensity (integrated along the spatial dimension) plotted vs column number. The plot shows multiple He-II lines, including 320.29 \AA, 303.78 \AA, 256.32 \AA, 243.03 \AA, and 237.33 \AA. The average Full Width at Half Maximum (FWHM) for these five lines is calculated to be 10.56 pixels which corresponds to a measured spectral resolution of 0.127 nm. Based on these lines and Zemax simulations, the instrument spectral range is estimated to be 16.1 nm to 33.8 nm along the 1504-pixel sensor direction.

\subsection{SXR Imaging Spectrometer Calibration}
A preliminary SXR imaging spectrometer calibration is planned to be performed using an Amptek MiniX2 source. This test will be performed in two configurations, first where the pinhole apertures are directly exposed MiniX2 source, and the second where the MiniX2 source is focused on a target, and the resulting X-ray fluorescence illuminates the pinhole apertures to get specific lines of interest. Fast readout will be implemented by on a selection of rows to obtain spectra of the source, by computing the energy of individual photons. Additionally, the SDO-EVE sounding rocket X55 spectrometer that was previously calibrated using NIST SURF will be used as a secondary radiometric standard to calibrate the Mini-X source output. Additional detailed post-flight calibration at NIST SURF is planned for the rocket-EVE instruments, including SEUSHI in 2027.

\section{Summary and Future Plan}\label{sec:Future_Work}
The design and development for the SEUSHI instrument for a flight onboard the SDO-EVE calibration sounding rocket was described in this paper. Preliminary environmental tests were also performed on the instrument to ensure that it can survive the launch and space environments. These include 3-axis vibration tests to validate mechanical design as well as a vacuum test to validate the thermal design of the instrument, particularly the sensor cooling. The instrument is ready for flight onboard the SDO-EVE underflight calibration rocket, planned to launch in August 2026. Further improvements of the instruments are proposed before it is developed as a satellite instrument. These include optimization of the filter thicknesses and materials based on further analysis to better detect the HOPE phenomenon. Additionally, the geometry (including the pinhole to detector distance and pinhole aperture size) of the instrument can also be further optimized to improve spatial resolution. The EUV spectrograph part of SEUSHI will be optimized, such as improving the SNR of the bright lines by having a wider and taller entrance slit. Additional calibration of the SEUSHI instrument is also planned at NIST SURF after the sounding rocket flight, which would include calibrating both the SXR imaging spectrometer and the EUV spectrograph. The SEUSHI instrument is complementary to other SXR instruments currently in development such as MOXSI on CubIXSS \citep{caspi_slitless_2016} and the Marshall Grazing Incidence X-Ray Spectrometer (MaGIXS, MaGIXS-2) sounding rocket instrument \citep{savage_first_2023}. SEUSHI provides simultaneous imaging and spectroscopy of the soft X-ray (SXR) and extreme ultraviolet (EUV) regimes, to characterize plasma properties during all phases of the flare including impulsive phase, gradual phase, coronal dimming (CME proxy), and EUV late phase. The sounding rocket flight of the SEUSHI instrument serves as a technology demonstration, paving the way for a flight aboard future satellite missions for enhanced space weather predictions.

\section{Acknowledgements} \label{sec:ack}
The authors would like to acknowledge the SDO-EVE Calibration NASA Contract NAS5-02140 for supporting the build and launch of the SEUSHI instrument. The authors would like to acknowledge the GOES EXIS and the SDO-EVE projects for providing the spare thin-foil filters, the diffraction grating, and the CMOS image sensor. The authors also acknowledge the University of Colorado Boulder's Department of Aerospace Engineering Sciences vibration table facility for preliminary vibration testing of the SEUSHI instrument. 

% \section{Data Source} \label{sec:datasource}
% After launch of the SEUSHI instrument the data as well as analysis code will be available publicly via LASPs SDO EVE website. 

\appendix
\section{SXR Imaging Spectrometer Design Optimization}
\label{sec:appendixA}
The SXR imaging spectrometer has a number of design parameters (pinhole to sensor distance, pinhole aperture, sensor pixel size, filter material and thickness) that effect the performance of the instrument in terms of spatial resolution, energy resolution and energy range. This section describes some of these design tradeoffs. 

\subsection{Spatial Resolution Optimization}

\begin{figure}[!htbp]
    \centering
    \includegraphics[width=\linewidth]{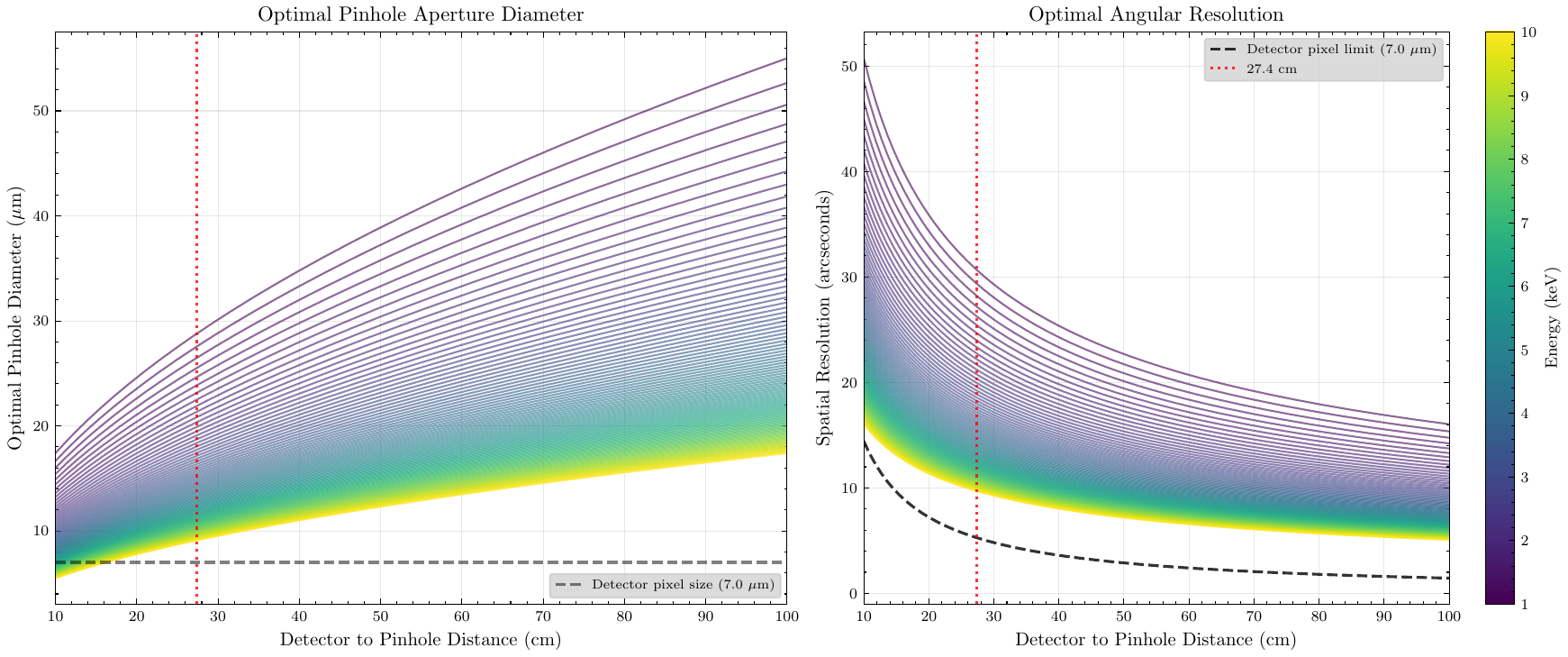}
    \caption{Left: Optimal Pinhole aperture diameter (microns) vs the sensor to pinhole distance (cm). Right: Spatial resolution (arcseconds) vs sensor to pinhole distance (cm). Both plots are shown for different energy levels from 1 to 10 keV. Dashed red vertical line indicates the current SEUSHI sensor to pinhole distance of 27.4 cm. }
    \label{fig:spatial_res}
\end{figure}

The spatial resolution of a pinhole camera is determined by two competing effects: geometric blur (arising from the finite size of the pinhole) and diffraction blur. The total blur is the quadrature sum of geometric and diffraction blur:
\begin{equation} \label{eq:spatial_res}
\delta_{\text{total}} = \sqrt{\delta_{\text{geom}}^2 + \delta_{\text{diff}}^2} = \sqrt{d^2 + \left(2.44 \frac{\lambda L}{d}\right)^2}
\end{equation}

where $\delta_{\text{geom}}$ is the geometric blur size, $\delta_{\text{diff}}$ is the diffraction blur size (the Airy disk diameter), $d$ is the pinhole aperture diameter,  $\lambda$ is the wavelength, and $L$ is the sensor-to-pinhole distance. To find the optimal pinhole diameter that minimizes the total blur, we differentiate Equation \ref{eq:spatial_res} with respect to $d$ and set the derivative to zero, which gives the optimal pinhole diameter as $\mathrm{d_{\text{opt}} = \sqrt{2.44 \lambda L}}$. Substituting the optimal diameter back into Equation \ref{eq:spatial_res}, the minimum total blur (best spatial resolution) for a given pinhole to sensor distance is given as $\mathrm{\delta_{\text{min}} = \sqrt{4.88 \lambda L}}$. The corresponding best angular resolution (in radians) is obtained by dividing the linear blur by the distance, and is given as $\mathrm{\theta_{\text{min}} = \sqrt{\frac{4.88 \lambda}{L}}}$. Figure \ref{fig:spatial_res} shows the variation of the optimal pinhole aperture diameter vs the sensor to pinhole distance in the left panel. The right panel shows the corresponding optimal spatial resolution (in arcseconds) with respect to the sensor to pinhole distance. The resolution limit due to the sensor pixel size is also shown as a dashed gray curve (corresponding to a 7 $\mu m$ pixel varies with distance, with a factor of 2 added for the Nyquist criterion). Since the optimal pinhole aperture diameter and corresponding optimal spatial resolution also have a energy dependence, these are plotted over the SXR energy range from 1 to 10 keV. Thus, it can be seen that increasing the distance beyond 40 cm gives diminishing returns to the spatial resolution. The pinhole aperture diameter for this length is in the range of 10-25 microns. The pinhole aperture diameter also effects the total signal that is produced by the sensor which is important to ensure sufficient signal to noise and is discussed further in the next subsection. Figure \ref{fig:spatial_res} plots also show a vertical dashed red line that indicates the current selected sensor to pinhole distance of 27.4 cm.

\subsection{Aperture Size and Signal Calculation}

\begin{figure}[!htpb]
    \centering
    \includegraphics[width=\linewidth]{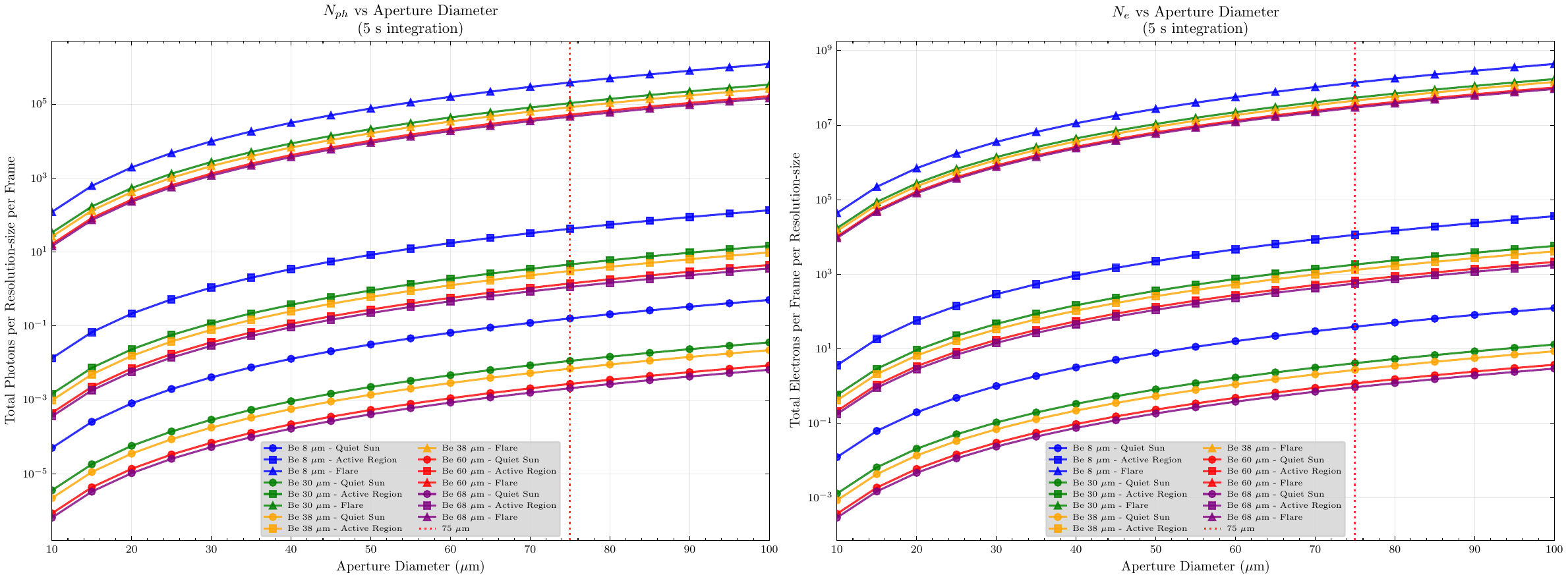}
    \caption{The left panel shows signal (photon per resolution-size per frame) vs pinhole aperture diameter for different Be filter thicknesses (8 $\mathrm{\mu}$m in Blue, 30 $\mathrm{\mu}$m in Green, 38 $\mathrm{\mu}$m in Yellow, 60 $\mathrm{\mu}$m in Red, and 68 $\mathrm{\mu}$m in Purple. These are quiescent sun (QS, denoted by circle markers), Active Region (AR, denoted by square markers), and Flare (FL, denoted by triangle markers) model DEM spectra generated from CHIANTI. The right panel shows a similar plot but with signal in terms of electrons per pixel per resolution-size per frame. Vertical dashed red lines in both plots indicate the 75 $\mathrm{\mu}m$ aperture size, which is the currently used aperture size for the SEUSHI instrument.}
    \label{fig:fluxdesign}
\end{figure}

The pinhole aperture diameter directly affects both the spatial resolution and the signal throughput. Additionally, the filter material and thickness (which is needed to block lower energy photons) also effect the total signal. The number of detected photons per resolution-size (i.e. aperture size on sensor) per frame is denoted by $N_{ph}$, and is computed as:
\begin{equation}
N_{ph} =
\frac{\pi d^2}{4}
\left(\frac{d}{L}\right)^2
\Delta t
\int_{\lambda}
L_\lambda(\lambda)\,
T_{\mathrm{Be}}(\lambda)\,
T_{\mathrm{SiO}}(\lambda)\,
\left[1-e^{-\mu_{\mathrm{Si}}(\lambda)t_{\mathrm{Si}}}\right]
\, d\lambda
\end{equation}
where, $d$ is the pinhole aperture diameter, $L$ is the pinhole-sensor distance, $\Delta t$ is the integration time per frame, $L_\lambda(\lambda)$ is the modeled Solar spectral radiance (photons cm$^{-2}$ sr$^{-1}$ s$^{-1}$ \AA$^{-1}$), $T_{\mathrm{Be}}(\lambda)$ is the transmission of the beryllium filter at wavelength $\lambda$, $T_{\mathrm{SiO}}(\lambda)$ is the transmission of the SiO layer on the sensor at wavelength $\lambda$, $\mu_{\mathrm{Si}}(\lambda)$ is the mass attenuation coefficient of silicon at wavelength $\lambda$, $t_{\mathrm{Si}}$ is the active silicon thickness of the sensor. To convert from number of detected photons to the actual signal generated by multiply by the energy of each photon, and divide by the energy required to generate one electron--hole pair in silicon, to get number of electron per resolution-size per frame ($N_e$):
\begin{equation}
N_e =
\frac{\pi d^2}{4}
\left(\frac{d}{L}\right)^2
\Delta t
\int_{\lambda}
L_\lambda(\lambda)\,
T_{\mathrm{Be}}(\lambda)\,
T_{\mathrm{SiO}}(\lambda)\,
\left[1-e^{-\mu_{\mathrm{Si}}(\lambda)t_{\mathrm{Si}}}\right]\,
\frac{hc}{\lambda w_{\mathrm{Si}}}
\, d\lambda
\end{equation}
where, $h$ is planck constant, $c$ is the speed of light, $\lambda$ is the photon wavelength, and $w_{\mathrm{Si}}$ is the mean energy required to produce one electron--hole pair in silicon ($\approx 3.65$ eV). Figure \ref{fig:fluxdesign} shows the variation of total signal vs pinhole aperture diameter for different Be filter thicknesses. The left panel shows $N_{ph}$ and the right panel shows $N_e$. Both the panels are shown for different Be filter thicknesses (8 $\mu$m, 30 $\mu$m, 38 $\mu$m, 60 $\mu$m, and 68 $\mu$m). These plots are shown for three different spectra generated using CHIANTI Differential Emission Measure (DEM) models for quiescent sun (QS), Active Region (AR), and Flare (FL). The QS DEM corresponds to A6 XRS-B irradiance, AR corresponds to C4, and FL corresponds to X4. Both the plots show signal per resolution-size per frame with an integration time of 5 seconds per frame. The right plot shows that the number of electrons per resolution size (which is aperture area on sensor, approximately 90 pixels) is less than $\mathrm{10^4~e^-}$) for 75 micron aperture size for all Be filter thicknesses for AR spectra. Thus, for this aperture size, the signal per pixel would be well below the full-well for active region spectra. Since the rocket flight is planned not to launch during the flare (for SDO-EVE calibration), the 75 micron aperture size is chosen to optimize for measurements of the active region spectra.

\subsection{Photon-counting Optimization}
\begin{figure}[!htpb]
    \centering
    \includegraphics[width=\linewidth]{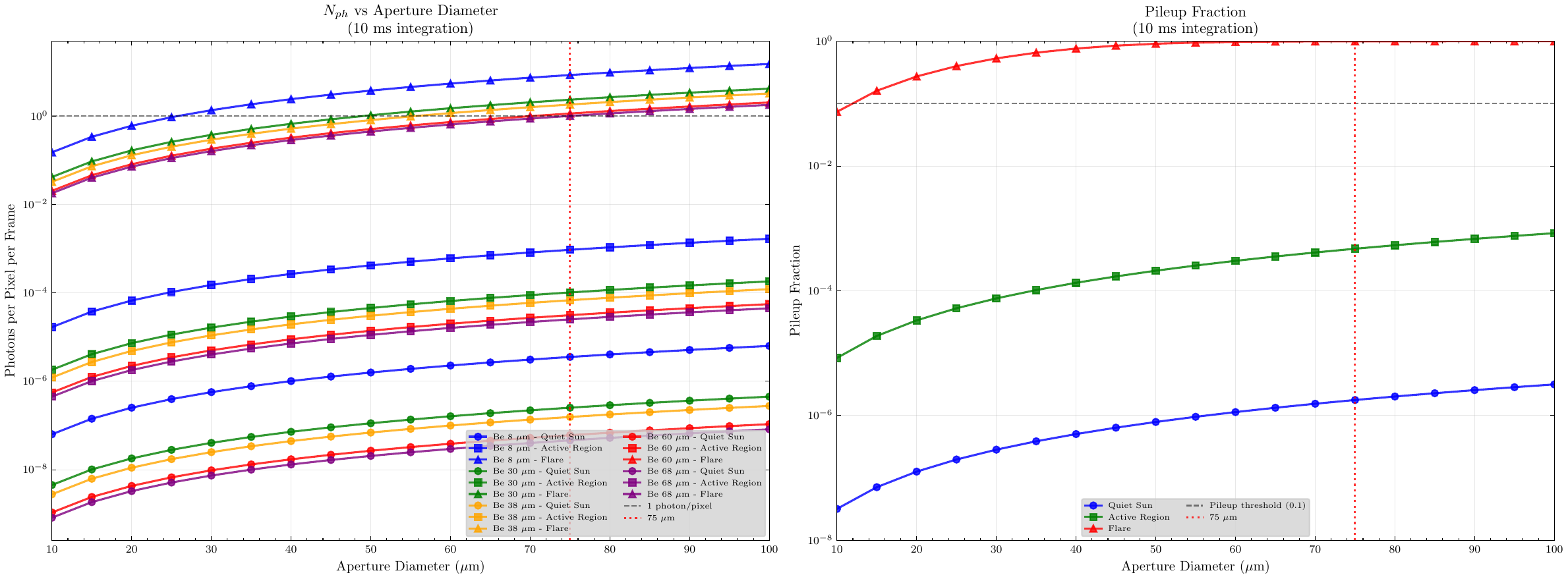}
    \caption{Left panel shows signal (photons per pixel per frame) vs aperture diameter for fast readout rate (100 Hz) to perform photon-counting for determining photon energy. The plot is shown for different Be filter thicknesses for different CHIANTI DEM model spectra including quiescent sun (QS), Active Region (AR), and Flare (FL). Right panel shows pile-up fraction vs aperture diameter for different DEM spectra and 8 micron Be filter thickness. Pileup threshold (black horizontal line) is shown as 0.1, indicating the upper limit level for photon-counting. Vertical dashed red lines in both plots indicate the 75 $\mathrm{\mu}m$ aperture size, which is the currently used aperture size for the SEUSHI instrument.}
    \label{fig:photoncounting}
\end{figure}

\begin{figure}
    \centering
    \includegraphics[width=0.92\linewidth]{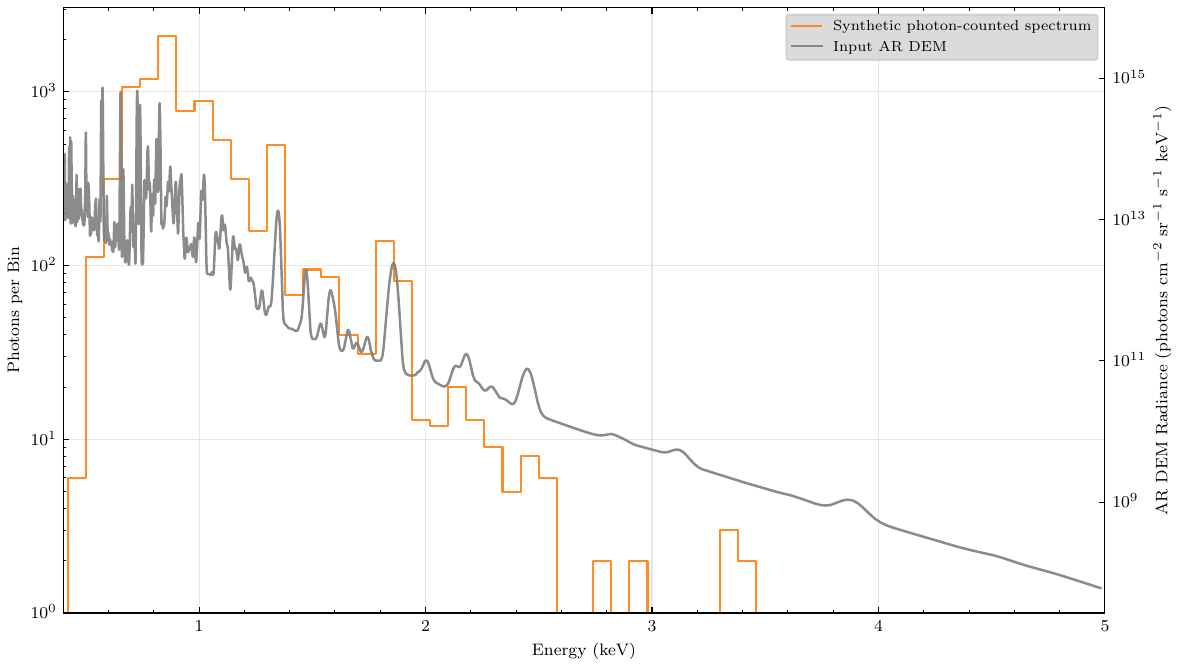}
    \caption{Synthetic photon-counting spectra for an active region generated over a duration of 5 minutes (typical observation window for the sounding rocket test flight). The gray line shows the active region DEM spectra obtained from CHIANTI. The orange line shows the synthetic spectra obtained by measuring charge generated by individual X-ray photons to estimate photon energy. The bin width (energy resolution) is 80 eV which calculated from the 3-sigma noise level of the instrument.}
    \label{fig:synt_spectra}
\end{figure}

In order to perform photon-counting, it must be ensured that the pileup fraction (number of photons hitting the same pixel within an frame integration time is $<<$ 1). This would allow estimating the energy of a photon by calculating the number of electron-hole pairs created by each photon. Assuming photon arrivals follow Poisson statistics with mean $r$ photons per frame per pixel, the fraction of events affected by pile-up is given by:
\begin{equation}
f_{\mathrm{pile}} = \frac{P(k \ge 2)}{P(k\ge1)}
= \frac{1 - e^{-r} - \lambda e^{-r}}{1 - e^{-r}},
\end{equation}
where $P(k) = \frac{r^k e^{-r}}{k!}$ is the probability of detecting k photons per pixel per frame, $P(0)=e^{-r}$ is the probability of detecting zero photons per pixel per frame, and $P(1)=re^{-r}$ is the probability of detecting 1 photon per pixel per frame. This expression represents the conditional probability that two or more photons arrive given that at least one photon was detected. The left panel of Figure \ref{fig:photoncounting} shows the signal per pixel vs aperture diameter for different thickness Be filter thicknesses with an frame integration time of 10 millisconds (100 Hz). The plot is shown for different model spectra (including quiescent sun (QS), Active Region (AR), and Flare (FL)). Note that this is per pixel and not per resolution-size like Figure \ref{fig:fluxdesign}. The right panel of Figure \ref{fig:photoncounting} shows the pileup fraction as a function of aperture diameter for different model spectra for the 8 micron Be filter that is used for photon-counting. The pile-up fraction threshold (shown by the black horizontal line) is set as 0.1 to ensure photon counting is feasible, by ensuring less than 10\% of events have multiple photon hits in a single frame.
For the rocket instrument, an analysis was conducted to obtain the synthetic spectra that would be observed during the 5-minute observation window. Figure \ref{fig:synt_spectra} shows the synthetic spectra simulated spectra (orange) overlayed over the active region DEM (gray) obtained from CHIANTI. This simulation is performed by taking the input spectra and convolving with the instrument response function. The signal is summed over a region of interest, which is the number of pixels illuminated by a typical 1 arcminute active region. Since the rocket instrument consists of six pinhole apertures out of which three images consists of rows where fast integration time is implemented, the plot shows the signal estimates summed over the three images. The measured spectra is binned according to the instrument resolution of 80 eV, which is discussed in Section \ref{sec:snr}.

Overall, from Figure \ref{fig:spatial_res} and Figure \ref{fig:fluxdesign}, it can be seen that increasing the aperture size degrades the spatial resolution but improves the total signal. For the SEUSHI instrument on the sounding rocket platform a pinhole to sensor distance of 274 mm was chosen with an aperture size of 75 microns. Although, this is not the optimum pinhole diameter for spatial resolution, it allows sufficient signal to enable photon counting to build up spectra of an active region during the sounding rocket flight. A future version of the SEUSHI instrument on a satellite platform can have longer pinhole to sensor distance and different aperture sizes for flare and active region observations. This would allow higher spatial resolution for characterizing the HOPE features, and higher signal for building up the spectra for active regions. 

\subsection{Signal to Noise Ratio Estimates}\label{sec:snr}
The calculation performed for estimating the Signal to Noise Ratio (SNR) for the SXR images, SXR fast-readout photon counting rows, and the EUV spectrograph are described in this section. The total noise per pixel (e.g. \cite{photontransfer}), denoted by $\sigma_{\mathrm{total}}$ is given as :
\begin{equation}\label{eq:noise}
\sigma_{\mathrm{total}} = \sqrt{ \sigma_{\mathrm{arrival}}^2 + \sigma_{\mathrm{fano}}^2 + \sigma_{\mathrm{dark}}^2 + \sigma_{\mathrm{read}}^2}
\end{equation}
The different noise sources in equation \ref{eq:noise} are arrival noise, fano noise, dark noise and read noise. Arrival noise,  $\sigma_{\mathrm{arrival}} = \sqrt{N_{ph}}\cdot (E_{ph}/w_{Si})$, is the photon arrival noise associated with Poisson statistics (also referred to as Shot noise). This depends on number of photons per pixel ($N_{ph}$), the energy of the photon ($E_{ph}$), and the energy required to generate one e-h pair in silicon (denoted by $w_{Si}$ which is approximately 3.65 eV). The Fano noise is $\sigma_{\mathrm{fano}} = \sqrt{F \, N_{e^-}}$ and is due to the fluctuations in charge generation, which depends on the Fano Factor (F, typically 0.1 for Si) and the number of electrons per pixel ($N_{e^-}$). The dark noise is given as $\sigma_{\rm dark} = \sqrt{I_d\, t}$, and depends on the dark current ($I_d$) and the integration time (t). According to the CIS115 datasheet the dark current is 20 $e^-$/pixel/second at 19.85$^\circ$C (293 K) and haves by a temperature reduction of 5.5 $^\circ$C.  
Thus, to achieve a dark current of 1~$e^-$/pixel/s, a temperature of approximately -6 $^\circ$C is required. The SEUSHI instrument is planned to be operated at -10$^\circ$C to ensure sufficient margin. \textbf{Thus, for the SNR analysis in this section the dark current number used corresponds to a temperature of -10$^\circ$C.} The readout noise is a property of the sensor and electronics, and is 5~$e^-$ per read for the CIS115 sensor. Table \ref{tab:noise_budget} summarizes the signal and noise calculations for different measurement scenarios of the SEUSHI instrument. 

\begin{deluxetable*}{lccccccc}[!htpb]
\tablecaption{Signal and noise budget comparison for SEUSHI in SXR imaging, SXR spectroscopy, and EUV spectroscopy observing modes.\label{tab:noise_budget}}
\tablewidth{0pt}
\tablehead{
\colhead{} &
\multicolumn{4}{c}{\textbf{Imaging}} &
\multicolumn{2}{c}{\textbf{Spectroscopy}} &
\colhead{\textbf{EUV}} \\
\colhead{} &
\colhead{\textbf{8 $\mu$m}} &
\colhead{\textbf{8 $\mu$m}} &
\colhead{\textbf{38 $\mu$m}} &
\colhead{\textbf{38 $\mu$m}} &
\colhead{\textbf{SXR}} &
\colhead{\textbf{SXR}} &
\colhead{\textbf{193 $\mathrm{\AA}$}} \\
\colhead{Measurement Scenario} &
\colhead{AR} &
\colhead{M1 flare} &
\colhead{AR} &
\colhead{M1 flare} &
\colhead{1 pixel} &
\colhead{4 pixels} &
\colhead{Line}
}
\startdata
Integration time ($t$) & $5$ s & $5$ s & $5$ s & $5$ s & $10$ ms & $10$ ms & $5$ s \\
Signal ($N_{\rm ph}$, photons pixel$^{-1}$) & $4.73\times10^{-1}$ & $1.07\times10^{2}$ & $3.42\times10^{-2}$ & $2.29\times10^{1}$ & $1$ & $1$ & $2.42$ \\
Signal ($N_e$, e$^{-}$ pixel$^{-1}$) & $1.27\times10^{2}$ & $3.85\times10^{4}$ & $1.45\times10^{1}$ & $1.27\times10^{4}$ & $5.48\times10^{2}$ & $1.37\times10^{2}$ & $4.26\times10^{1}$ \\
Signal (DN pixel$^{-1}$) & $2.54\times10^{2}$ & $7.70\times10^{4}$ & $2.90\times10^{1}$ & $2.53\times10^{4}$ & $1.10\times10^{3}$ & $2.74\times10^{2}$ & $8.51\times10^{1}$ \\
Photon arrival noise & $1.88\times10^{2}$ & $2.83\times10^{3}$ & $5.07\times10^{1}$ & $1.31\times10^{3}$ & N.A. & N.A. & $2.74\times10^{1}$ \\
Fano noise & $3.56$ & $6.21\times10^{1}$ & $1.20$ & $3.56\times10^{1}$ & $7.40$ & $3.70$ & $2.06$ \\
Dark noise & $2.24$ & $2.24$ & $2.24$ & $2.24$ & $0.1$ & $0.1$ & $2.24$ \\
Read noise & $6$ & $6$ & $6$ & $6$ & $6$ & $6$ & $6$ \\
Total noise & $1.89\times10^{2}$ & $2.83\times10^{3}$ & $5.11\times10^{1}$ & $1.31\times10^{3}$ & $9.53$ & $7.05$ & $2.82\times10^{1}$ \\
SNR (per pixel) & $6.73\times10^{-1}$ & $1.36\times10^{1}$ & $2.84\times10^{-1}$ & $9.65$ & $5.75\times10^{1}$ & $1.94\times10^{1}$ & $1.51$ \\
SNR (per resolution element) & $6.39$ & $1.29\times10^{2}$ & $2.69$ & $9.16\times10^{1}$ & --- & --- & --- \\
SNR (res. el., 60 s avg.) & $2.21\times10^{1}$ & $4.47\times10^{2}$ & $9.33$ & $3.17\times10^{2}$ & --- & --- & --- \\
SNR (pixel neighborhood) & --- & --- & --- & --- & --- & $3.89\times10^{1}$ & --- \\
SNR (per column) & --- & --- & --- & --- & --- & --- & $3.27\times10^{1}$ \\
SNR (per line) & --- & --- & --- & --- & --- & --- & $5.67\times10^{1}$ \\
SNR (line, 60 s avg.) & --- & --- & --- & --- & --- & --- & $1.96\times10^{2}$ \\
\enddata
\tablenotetext{}{Abbreviations: AR = active region; DN = data number; $N_{\rm ph}$ = detected photons per pixel; $N_e$ = generated electrons per pixel; SNR = signal-to-noise ratio; res.\ el. = resolution element which is approximately 90 pixels.}
\end{deluxetable*}

\textbf{For the SXR images (first four columns of table \ref{tab:noise_budget}) the SNR estimates are done for an active region (since that will likely be observed during the rocket flight) and a M1 flare (since that is the requirement according to the Science Traceability Matrix (table \ref{tab:stm})). Similar to the left plot in Figure \ref{fig:fluxdesign}, the photon flux per pixel was calculated using CHIANTI DEM spectra for both active region and a M1 flare. The number of electrons per pixel was also calculated using the silicon detector gain (similar to the right plot in Figure \ref{fig:fluxdesign}). Note that figure \ref{fig:fluxdesign} shows signal per resolution element, whereas table \ref{tab:noise_budget} lists signal per pixel in the second row. Assuming that the photons are distributed uniformly over a 75~$\mu$m diameter resolution element (area $\approx$ 4418~$\mu$m$^2$) and sampled by 7~$\mu$m  pixels (area 49 ~$\mu$m$^2$), the resolution element spans approximately 90 pixels. The analysis is done for both 8~$\mu$m  and 38~$\mu$m  Be filter channels of the SEUSHI instrument. The table shows the modeled signal, different noise sources as well as the SNR. It can be observed that the noise is highly shot noise dominated and with the resulting signal-to-noise ratio per pixel for M1 flare and 8~$\mu$m  filter being  $\mathrm{SNR}_{\mathrm{pix}}$ $\approx$ 13.6. When summed over the full resolution element (90 pixels), the total signal-to-noise ratio improves as $\sqrt{90}$, yielding $\mathrm{SNR}_{\mathrm{res}}$ $\approx$ 129. For the active region imaging during the sounding rocket flight the SNR per resolution element is 6.4 for the 8~$\mu$m  filter channel and 2.7 for the 38~$\mu$m filter channel. In order to improve the SNR, multiple images will be averaged during the sounding rocket flight. The SNR (res. el., 60 s avg) row of table \ref{tab:noise_budget} shows that for a 60 second average the SNR improves to 22 for the 8~$\mu$m filter channel and 9.3 for the 38~$\mu$m filter channel.}

For SXR spectroscopy, the fifth column of the table shows the calculation for the photon-counting scenario, with the example of a single 2 keV photon hitting a single pixel. Previous work (for example \cite{photontransfer, heymes_development_2020}) has shown that for a CMOS sensor, a typical X-ray photon generates electron-hole pairs in a neighborhood of four pixels. The SNR for this scenario is calculated in the sixth column of the table. For a 2~keV photon incident on a silicon detector, the number of generated electrons in the neighborhood is $N_{e^-} = E_\mathrm{ph}/w_{\mathrm{Si}} \approx 2000/3.65 \approx 548~e^-$, which implies $137~e^-$ per pixel. For the photon-counting measurement scenario, arrival noise is 0, number of photons hitting each pixel is $<<$ 1 during an integration time. The total noise is computed as approximately 7 electrons/pixel, which corresponds to a full-width half max energy resolution of 60 eV, and a 3-sigma resolution of 77 eV at 2 keV. The fifth and sixth columns shows the SNR for a single-pixel interaction is approximately 57.6 and for a interaction with a neighborhood of 4 pixels is 38.9. The last column shows the SNR estimation for the EUV spectrograph, at a line feature of 193 $\mathrm{\AA}$ which is of particular interest for the coronal dimming studies. Using signal estimates from figure \ref{fig:euv_signal}, the SNR per-pixel is computed to be 1.51, which is then multiplied by $\mathrm{\sqrt{N_{col}\cdot W_{line}}}$, where $N_{col}$ is the number of pixels in a column (approximately 470) and $W_{line}$ is the width of the 193$\mathrm{\AA}$ line (approximately 3), to get SNR for the 193$\mathrm{\AA}$ line which is approximately 196. A future version of the SEUSHI instrument can utilize a wider and taller slit to further improve the SNR. 
\bibliography{references_zotero, references_custom}{}
\bibliographystyle{aasjournalv7}
\end{document}